\documentclass{nature}
\usepackage{xargs}                      
\usepackage[pdftex,dvipsnames]{xcolor}  
\usepackage[colorinlistoftodos,prependcaption,textsize=tiny]{todonotes}
\newcommandx{\unsure}[2][1=]{\todo[linecolor=red,backgroundcolor=red!25,bordercolor=red,#1]{#2}}
\newcommandx{\change}[2][1=]{\todo[linecolor=blue,backgroundcolor=blue!25,bordercolor=blue,#1]{#2}}
\newcommandx{\info}[2][1=]{\todo[linecolor=OliveGreen,backgroundcolor=OliveGreen!25,bordercolor=OliveGreen,#1]{#2}}
\newcommandx{\improvement}[2][1=]{\todo[linecolor=Plum,backgroundcolor=Plum!25,bordercolor=Plum,#1]{#2}}
\newcommandx{\thiswillnotshow}[2][1=]{\todo[disable,#1]{#2}}
\usepackage[normalem]{ulem}
\usepackage{framed} 
\usepackage{float}
\restylefloat{table}
\usepackage{color}
\usepackage{amssymb}
\usepackage{mathrsfs}
\usepackage{epstopdf}
\usepackage{makeidx}
\usepackage{bm}
\usepackage{epsf,amsfonts,amsmath}
\usepackage{amssymb,amsmath}
\usepackage{appendix}
\usepackage{graphicx}
\usepackage{multirow}
\usepackage{booktabs}
\usepackage{array, makecell}
\usepackage{url}
\usepackage{soul}
\usepackage[version-1-compatibility]{siunitx}
\usepackage[percent]{overpic}
\usepackage{siunitx,booktabs,tabularx,ragged2e}

\usepackage{lineno}

\newcommand{\figref}[1]{Fig.~\ref{#1}}
\newcommand{\tabref}[1]{Table~\ref{#1}}

\newcommand{\sobol}{Sobol'}

\newcommand{\opal}{\textsc{OPAL}}

\newcommand{\psims}{physics simulations}
\newcommand{\psim}{physics simulation}

\newcommand{\moo}{GA}

\makeatletter
\let\saved@includegraphics\includegraphics
\AtBeginDocument{\let\includegraphics\saved@includegraphics}
\renewenvironment*{figure}{\@float{figure}}{\end@float}
\makeatother

\title{Machine Learning for Orders of Magnitude Speedup in Multi-Objective Optimization of Particle Accelerator Systems}

\author{Auralee Edelen$^{1}$, Nicole Neveu$^{1}$,  Matthias Frey$^2$, Yannick Huber$^{2}$, Christopher Mayes$^{1}$,
Andreas Adelmann$^{2}$}

\begin{document}
\maketitle 
\begin{affiliations}
 \item SLAC National Laboratory, Menlo Park, CA, USA
 \item Paul Scherrer Institut, Villigen, Switzerland 
\end{affiliations}

\begin{abstract}




 In addition, even  simulations with the main expected physics effects included often deviate substantially from the observed machine behavior.

High-fidelity \psims\ are powerful tools in the design and optimization of charged particle accelerators. However, the computational burden of these simulations often limits their use in practice for design optimization and experiment planning. It also precludes their use as online models tied directly to accelerator operation. We introduce an approach based on machine learning to create nonlinear, fast-executing surrogate models that are informed by a sparse sampling of the \psim. The models are $\mathcal{O}(10^{6})$--$\mathcal{O}(10^{7})$ times more computationally efficient to execute. We also demonstrate that these models can be reliably used with multi-objective optimization to obtain orders-of-magnitude speedup in initial design studies and experiment planning. For example, we required 132 times fewer simulation evaluations to obtain an equivalent solution for our main test case, and initial studies suggest that between 330--550 times fewer simulation evaluations are needed when using an iterative retraining process. Our approach enables new ways for high-fidelity particle accelerator simulations to be used, at comparatively little computational cost.

\end{abstract}

\section*{Introduction}
Physics simulations are essential tools for the initial design of modern particle accelerator systems, as well as for the subsequent optimization of new operating configurations. However, there is generally a tradeoff between simulation speed and accuracy in terms of the represented physics effects. Standard codes for simulating accelerator systems can be computationally intensive to run, particularly when complex beam behavior must be taken into account (e.g.,\ instabilities, collective effects, beam self-fields). Exacerbating this computational burden, accelerator systems often consist of many components that can be used to accelerate and manipulate the beam (e.g.,\ accelerating cavities, bending and focusing magnets, collimators). Each of these components has controllable variables that can be independently adjusted to achieve specific beam characteristics. In many cases, the subtle interactions between all variables must be considered. Thus, modeling these systems from ``start-to-end'' (i.e., from the beginning of the accelerator to a final point of interest) is critical for obtaining realistic predictions. As a result, design and optimization studies for particle accelerator systems often require the use of thousands of cores at High Performance Computing (HPC) facilities. While in principle many large accelerator facilities have access to such resources, in practice this computational burden significantly hampers efforts to conduct comprehensive optimization studies.

Optimization studies are important in the initial design of particle accelerator systems, when many tradeoffs between possible setting combinations have to be explored. In practice, multi-objective optimization with genetic algorithms (GAs)~\cite{mitchellga,Backga} is frequently used for finding optimal setting combinations (see~\cite{bazarov,hofler13,Neveu:2013ues} for accelerator-specific examples). One advantage of using multi-objective optimization is that it enables one to examine optimal trade-offs between achievable beam parameters.\ This is done via examination of the estimated Pareto fronts, which delineate the limit at which one can no longer improve a particular parameter without negatively impacting another parameter. These trade-offs drive the selection of machine working points, which in turn guide the rest of the design process (such as selection of rf equipment with appropriate specifications). 

For accelerators that are already in operation, offline optimization is also used to aid in experiment planning and setup. This is especially the case for facilities that require frequent re-tuning of settings. For example, at free electron laser (FEL) facilities like the Linac Coherent Light Source (LCLS) and Swiss Free Electron Laser (SwissFEL), user requests for specific beam parameters need to be handled quickly and efficiently, and new configurations (e.g., novel FEL schemes) are often developed during limited blocks of time between the scheduled user experiments.

Even though high-fidelity \psims\ are often created as part of the initial design process for a new accelerator, they are often not fully utilized during machine operation (i.e.,\ as ``online models'') for on-the-fly optimization and control. Online models can be used in model-based control and can provide estimates of normally-inaccessible beam parameters. They can also be used to perform rapid analysis in the control room and to plan out new courses of action as goals during an experimental shift may change. In addition, online models can also be used to help identify when anomalous conditions have arisen (for example, if model predictions suddenly show a sharp increase in error). High-fidelity physics simulations are typically not used online due to their computational expense: the execution speed is often too slow to aid operation. Instead, online models tend to rely on greatly simplified representations of the machine physics (e.g., see  \cite{lcls_online, Pelaia:2013nva, virtual_accel}), and as a result trade accuracy for speed. 

In light of these limitations, improving the execution speed and scalability of particle accelerator simulations is an area that has 
seen considerable effort in recent years \cite{scidac1,scidac2}. Approaches to do this have focused on parallelization (e.g., see  \cite{Adelmann:2016buh})
and hardware-based acceleration of existing simulation codes (e.g., using GPUs) \cite{elegant,parmila}.  In a few exceptional cases, computationally expensive models have been used to aid live operation when on-site HPC resources are available~\cite{gpu_parmila, APS_online, cornell_online}. Improvements to underlying modeling algorithms, such as using the Lorentz boosted frame \cite{vay} and spectral solvers \cite{vay13,lehe16}, have also provided orders of magnitude increases in computation speed. All of these efforts are highly successful. However, it remains the case that the computational expense of these simulations prevents them from being fully utilized by the accelerator community.

Here we explore a different, but complementary, approach that immediately enables new capabilities in how these existing high-fidelity \psims\ can be used by the particle accelerator community. We show that one can create Machine Learning (ML) based surrogate models to obtain accurate, fast-executing representations of the relevant beam dynamics from a sparse sampling of the \psim\ of interest. In contrast to the \psim, the ML models can execute in fractions of a second on a laptop with comparable accuracy in predicting the resultant beam parameters. We also show that these models are useful for multi-objective optimization in two important ways: (1) they can accurately reproduce optimization results obtained from the physics simulation, meaning they can be reliably used in experiment planning and live optimization during accelerator operation, and (2) they can be used to substantially speed up the initial design process by eliminating the need to run an optimization algorithm entirely on the simulation.  In addition, although we do not address it in this work, these models can in principle be updated with machine measurements (e.g., see ~\cite{aedsurro3} for an initial example) to help improve model fidelity with respect to the real machine behavior.

We have used ML models in several previous instances to create fast-executing surrogates for computationally intensive accelerator simulations \cite{aedsurro1,FAST,aedsurro2,aedsurro3,aa1}. Here, we build on those works, but take a substantial step forward by evaluating such models for use in optimization (and multi-objective optimization in particular), which is one of the main desired use-cases for ML in particle accelerator applications \cite{mlwhitepaper, nnsurvey}. As such, this work represents an important contribution to the particle accelerator community. We also evaluate how many training samples are needed to obtain an accurate model when used for optimization and make a brief comparison between different classes of ML models.

We demonstrate the proposed approach considering two different types of accelerator systems: the injector at the Argonne Wakefield Accelerator (AWA) Facility \cite{awa} (a linear accelerator) and a high-intensity cyclotron proposed for the search for sterile neutrinos. The latter system is based on the Isotope Decay At Rest (IsoDAR) design, as detailed in \cite{PhysRevLett.109.141802}. The AWA injector has a simple layout that is very similar to that used by other accelerator facilities, whereas the cyclotron is a substantially more complex machine \cite{meiss}. These two cases were chosen specifically to show the generality of the proposed approach. For simulating both accelerators, the \opal\ simulation framework is used. OPAL is a parallel, particle-in-cell (PIC) code that handles nonlinear and collective beam effects (e.g., coherent synchrotron radiation, 3D space charge). 

It is common practice in the accelerator community to use GAs for multi-objective optimization, although alternatives exist, such as particle swarm optimization \cite{pso_1, pso_2}. Because of its ubiquity, we chose to run the popular NSGA-II~\cite{nsgaii} algorithm with the ML models as our standard for assessing their performance.  For brevity, in the subsequent text we refer to the OPAL simulation of the AWA and IsoDAR as the  ``\psim,'' and we refer to NSGA-II as the ``\moo".


\section*{Results} 

\subsection {Description of ML Approach and Validation Procedure}
\begin{figure}[ht!]
\centering
 {\includegraphics[width=0.6\textwidth]{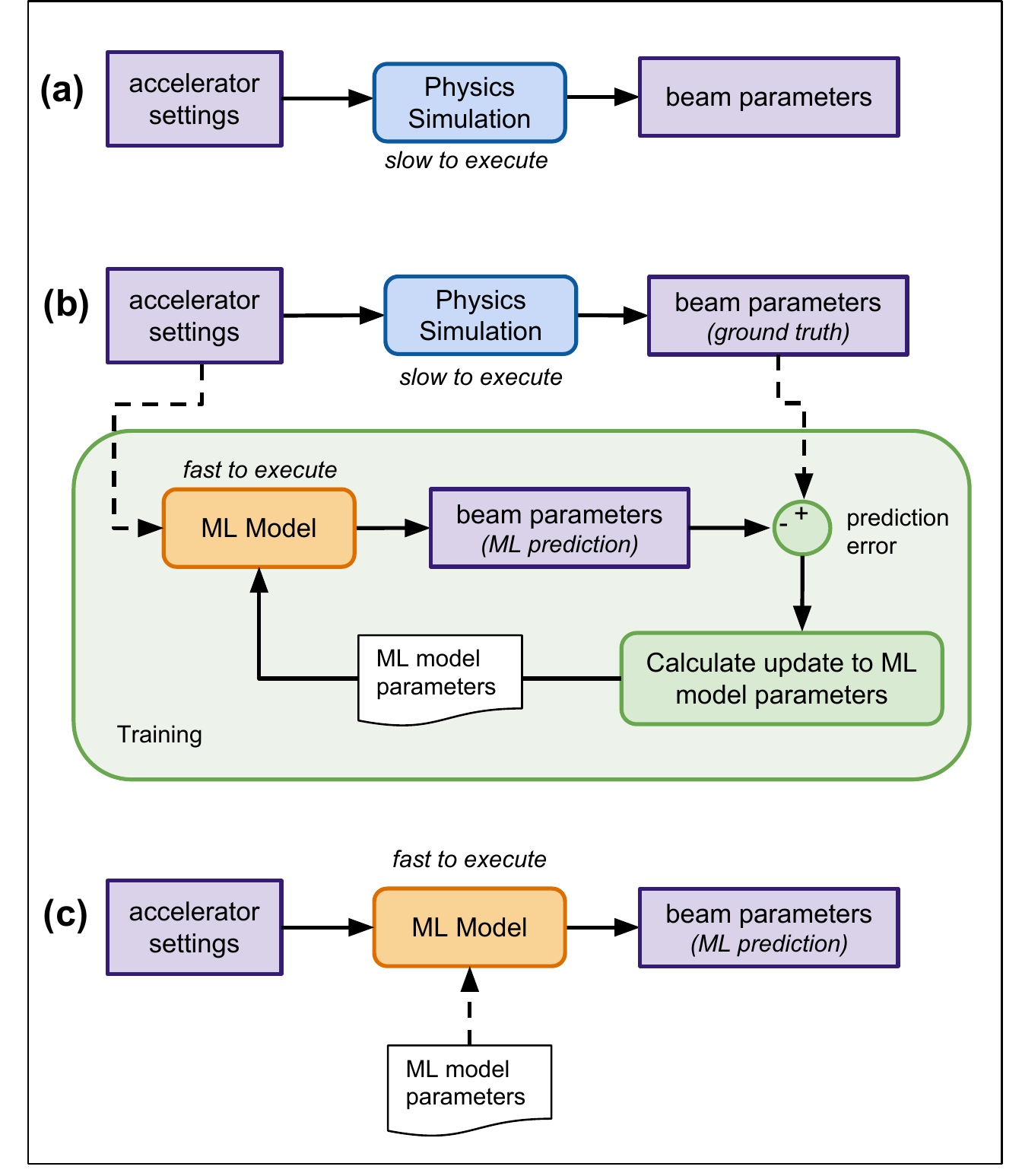}}
 \caption{ Initially, we have a computationally expensive \psim\ (a). We then use the \psim\ to generate a sparse set of training data for the ML model that covers a wide range of input settings. The ML model parameters are then optimized until the predictions of the beam parameters match those from the physics simulation (b). The result is a fast-executing representation of the \psim\ that can be used for optimization and online modeling (c).   \label{fig: surro-diag1}}
\end{figure}

\begin{figure}[ht!]
\centering
 {\includegraphics[width=0.9\textwidth]{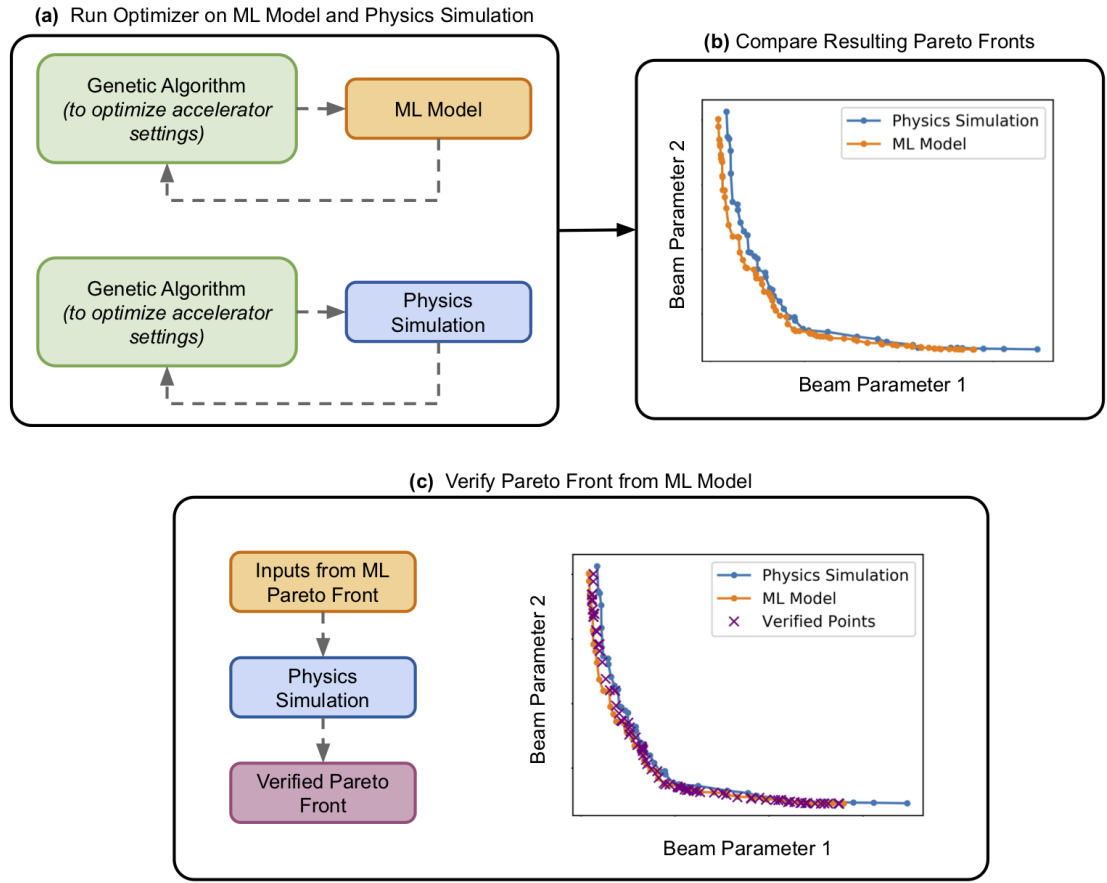}}
  \caption{Approach for assessing the reliability of ML-based model when used for optimization of beam parameters.
 	We run a \moo\ with the \psim\ to find accelerator settings (e.g., rf cavity phases, rf cavity gradients, solenoid strengths) that optimize the beam parameters. The optimization is repeated using the ML model instead of the \psim.  We then compare the estimated Pareto fronts for key beam parameters. Finally, we run the inputs corresponding to the estimated Pareto front predicted by the ML model through the simulation to verify the accuracy of the front. \label{fig: surro-diag2}}
\end{figure}

The general procedure for creating the ML surrogate models is shown in \figref{fig: surro-diag1}, and the procedure for using these models to improve the speed of optimization of particle accelerator systems is shown in \figref{fig: new_meth}. An ML model is trained on a sparse random sample of the accelerator input variables and the resulting beam parameters. The ML model can then be used as a fast-executing representation of the \psim. To assess the performance of the ML model when used with an optimization algorithm (see \figref{fig: surro-diag2}), we run a \moo\  with the  \psim\ to optimize settings (e.g., rf cavity phases, rf cavity gradients, solenoid strengths). We then run a  \moo\ with the ML model and compare the resultant estimated Pareto fronts. Good agreement between the estimated Pareto fronts indicates that the ML model can be used as an accurate replacement for the \psim\ in multi-objective optimization. We also take the input points that correspond to the estimated Pareto front from the ML model and run these through the physics simulation to verify the accuracy of the predictions.

\clearpage
\subsection {Validation of ML Surrogate Modeling Approach for Optimization}

We chose to assess this approach first with the AWA linear accelerator. Research at the AWA is focused on advanced accelerator concepts, which generally include efforts to improve control, diagnostic instrumentation, and components (e.g., accelerating structures) for future accelerators. 
Much effort is also dedicated to developing and testing beamline configurations that could be used for beam shaping~\cite{eex}, 
or future linear colliders~\cite{gai_power_jing_2012}. Often, the accelerator settings (e.g., focusing fields for all magnets, cavity phases, cavity accelerating gradients) are adjusted prior to each experiment to achieve custom beam characteristics (e.g.,  bunch length and transverse sizes). 
The accelerator also regularly operates at bunch charges where nonlinear effects are important (e.g., \SI{40}{nC}), and the cavity fields contain asymmetries. Overall, this results in a challenging optimization problem, and 3D PIC simulations are required to accurately predict the beam behavior. A fast-executing, accurate model of the machine could be useful for supporting the research program of the AWA. Taken together, these factors make the AWA a good test case.

\begin{figure}[ht!]
\centering
 {\includegraphics[width=0.7\textwidth]{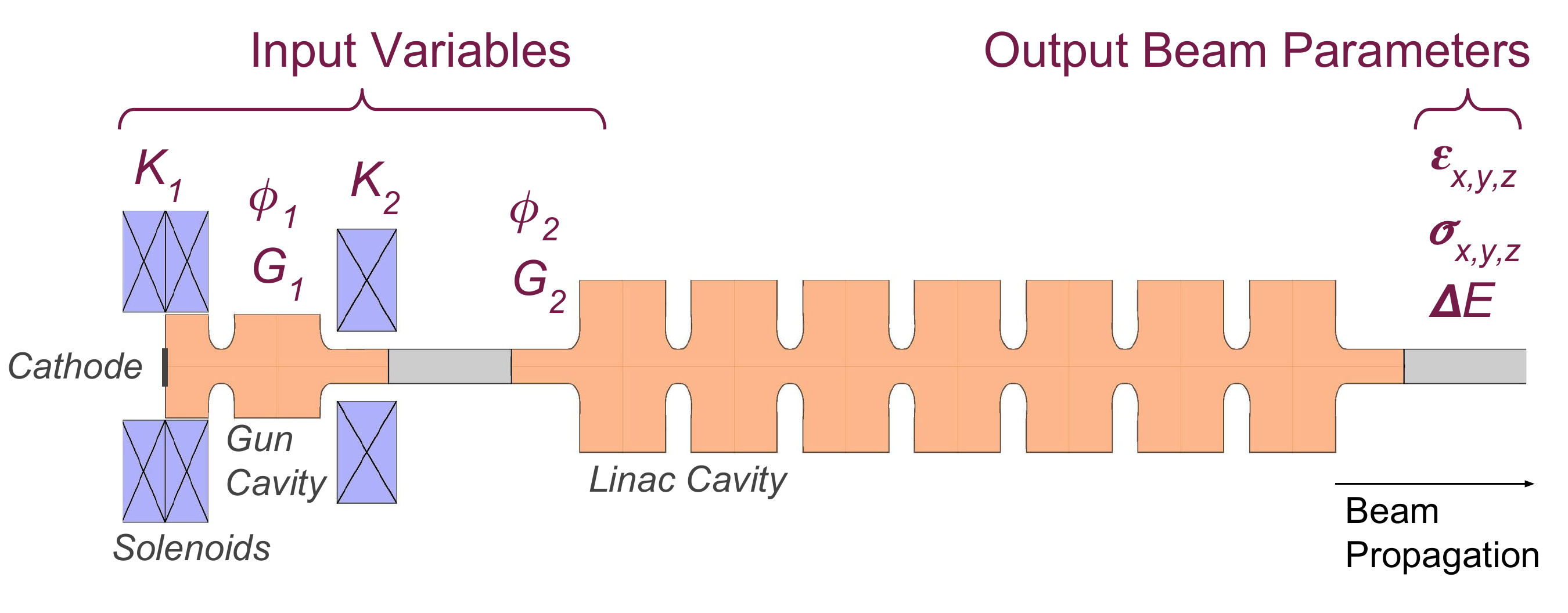}}
  \caption{Schematic of the AWA linac, together with the controllable accelerator settings and predicted beam parameters. The randomly-varied inputs include the injector rf phase $\phi_1$ and accelerating gradient $G_1$, the linac cavity rf phase $\phi_2$ and accelerating gradient $G_2$, and two solenoid strengths $K_1$ and $K_2$. The output electron beam parameters are the transverse spot sizes $\sigma_x$ and $\sigma_y$, the bunch length $\sigma_z$, the transverse projected emittance values $\varepsilon_x$ and $\varepsilon_y$, the longitudinal projected emittance $\varepsilon_z$, and the energy spread $\Delta$$E$. The input variable ranges are determined by typical operating ranges at the AWA and are shown in \tabref{tab: desparam}. We examined this setup for \SI{40}{nC} and \SI{1}{nC} bunch charges.}\label{fig:awa-1}
\end{figure}

We demonstrate the efficacy of the ML approach by training models on a sparse 
random sample of six adjustable input variables for the AWA and seven of the resultant beam parameters (see \figref{fig:awa-1}). Definitions of the output beam parameters can be found in \tabref{tab: beamparam} in the Appendix. The inputs were varied uniformly over a relevant operating range of the accelerator (see \tabref{tab: desparam}), and
 the same range of input variables was allowed for the \moo-based optimization of the beam parameters.
While the main focus was on an accelerator configuration with a bunch charge of  \SI{40}{nC}, we also examined a case with \SI{1}{nC} bunch charge (where nonlinear effects are less important).
Details on the data sets, training procedures, implementations of the ML models and the \moo, and the details of the \psims\ can be found in the Appendix. We first focus on Artificial Neural Networks (NNs) to demonstrate the technique, and later briefly compare the results with those obtained from Polynomial Chaos Expansion (PCE)~\cite{Smith2014,aa1} and Support Vector Regression (SVR) models.  

\begin{figure}[ht!]
	\centering
	{\includegraphics[width=0.9\textwidth]{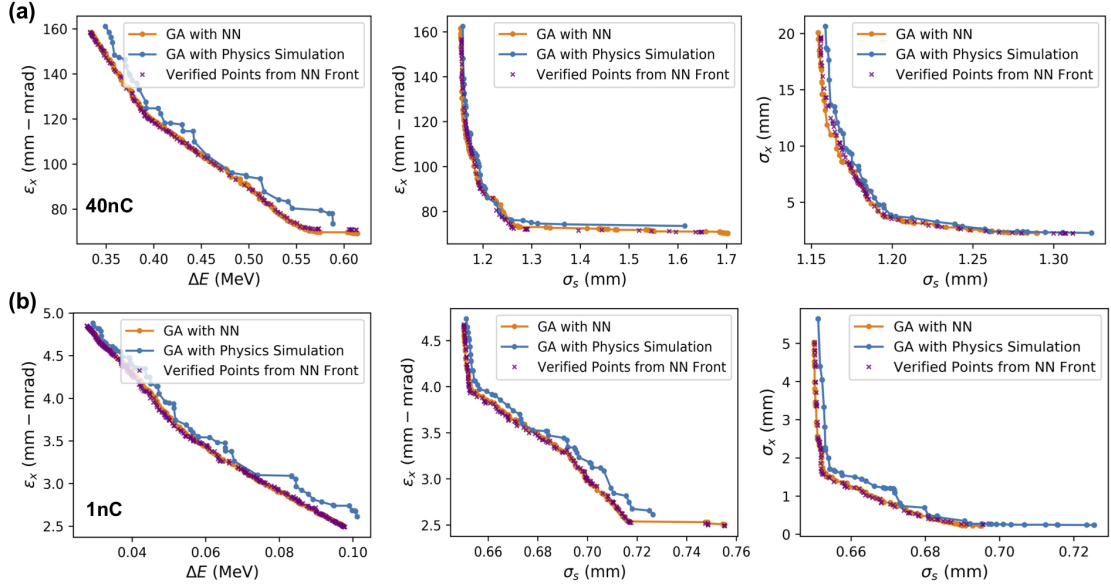}}
	\caption{Comparison between estimated Pareto fronts obtained from the NN and the physics simulation for three sets of beam parameters. We show results at the \SI{40}{nC} bunch charge (a) and at the \SI{1}{nC} bunch charge (b) configurations, and we find excellent agreement between the estimated Pareto fronts. In total seven beam parameters have been optimized, and we show examples of the 2D projections from this larger front for the parameters that are most critical for optimization of the AWA. The other projections show similar agreement.}  \label{fig: pareto_1nC40nC}
\end{figure}

The estimated Pareto fronts obtained using the NN models and the \psim\ are in good agreement (see \figref{fig: pareto_1nC40nC}). Only 500 random sample points were needed to train the NN in the \SI{40}{nC} case and still generate a set of Pareto fronts that is very close to those obtained with the \psim. In contrast, obtaining the same result with the \psim\ required just under 66,000 simulation evaluations. Furthermore, only a small amount of fine-tuning of the NN architecture was done in this case, as the initial topology and hyperparameters were chosen based on previous experience of the authors with similar types of injector modeling problems \cite{FAST,aedsurro1,aedsurro2,aedsurro3}. This highlights the generality of this approach for common kinds of accelerator components and hints at the possibility for doing transfer learning with the produced models (in which a pre-trained model can be applied to a new system with relatively little retraining). 

It is important to note that verifying the estimated Pareto front that was obtained with the NN model by running these points through the physics simulation does not verify that we have reached the actual Pareto front for the problem. Similarly, although we examined the convergence of the GA that we ran on the physics simulation to ensure it had converged to a stable solution, this is also not necessarily the true Pareto front for the problem. It is simply the one that the GA was able to converge to. Hence, in our assessment we only claim that the estimated Pareto front found with the NN model matches the estimated Pareto front obtained by running the GA on the physics simulation (i.e. which is our “ground truth” for this comparison). 

\begin{figure}[ht!]
\begin{minipage}[b]{1.0\linewidth}
\centering
  {\includegraphics[width=0.9\textwidth]{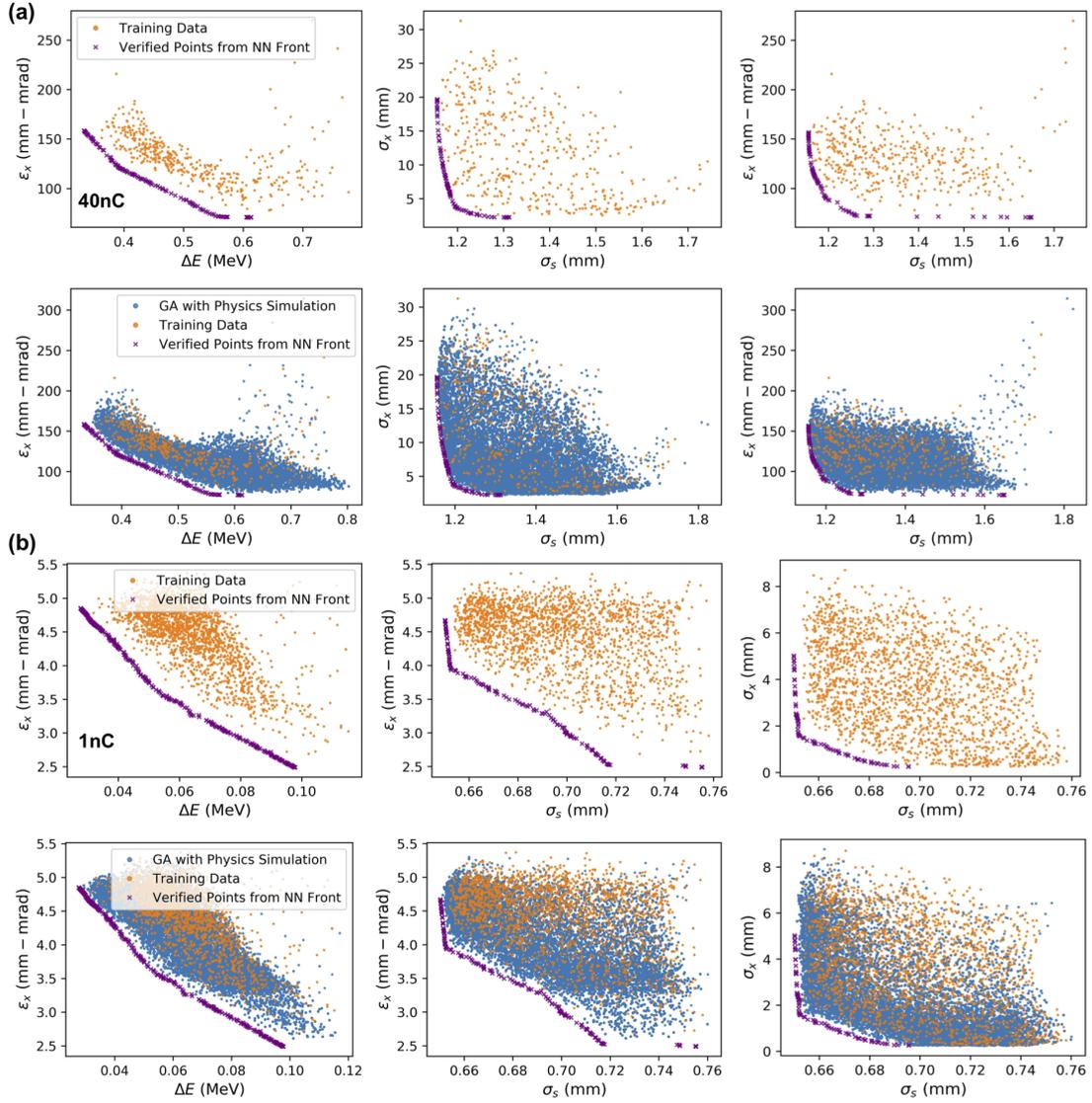}}

  \caption{Visualization of the improved solution found with optimizing on the NN model, as compared with both the training set and the points sampled with the GA that was run on the physics simulation. The position of the front indicates that the NN is able to interpolate in the input space to produce better combinations of output parameters than are observed in the training data.  We show the sampled points in the training set and the verified points from the estimated Pareto front from the NN for \SI{40}{nC} (a) and \SI{1}{nC} (b). For comparison, we also show the points sampled by the GA that was run on the physics simulation. \label{fig: example-par}}
  \end{minipage}
\end{figure}

In order to visualize the extent to which the NN is generalizing to new regions of the parameter space (as opposed to just learning the estimated Pareto front directly from the training data), we compared the training data with the final estimated Pareto fronts obtained with the NN (see~\figref{fig: example-par}). From this we infer that for some beam parameter combinations, the estimated Pareto front is in a region of the parameter space that is not sampled in the training data. This indicates that the NN is able to interpolate in the input parameter space (6 input dimensions, 7 output dimensions) to find optimal combinations of output beam parameters that are outside of the convex hull of those observed in the training data. In other words, these correspond to more optimal combinations of output beam parameters than were observed during training. Here ``more optimal'' means that for a given value of an output beam parameter that one would like to minimize, a solution was found where a competing output beam parameter that one also would like to  minimize is at a lower value than was obtained with the previous solution.

\clearpage
\subsection {Reducing Training Sample Size}

Producing training data using the \psim\ is computationally expensive. In light of this, one question which arises is how the accuracy of the ML model will change with the number of samples used in training. This is important for estimating the corresponding trade-off between computation time needed to generate the training set and the resultant model accuracy (the requirements for which may vary depending on the application).

\begin{figure}[h!]
	\centering
	{\includegraphics[width=0.9\textwidth]{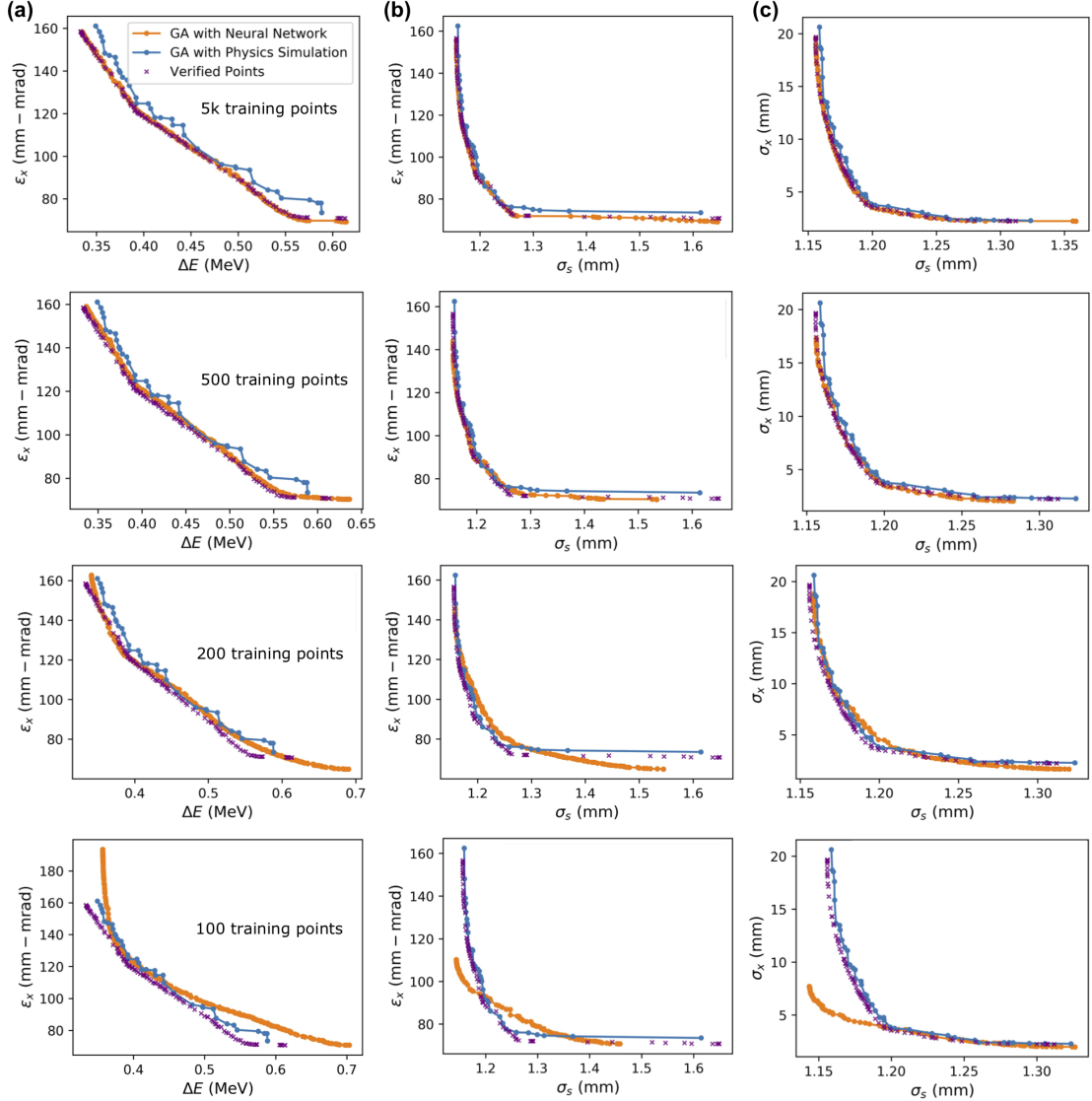}}
	\caption{Impact of training sample size on the quality of the estimated Pareto front solutions. We show a comparison between estimated Pareto fronts obtained from the NN for three sets parameters in the 40nC case: $\Delta$$E$ vs. $\varepsilon_x$ (a), $\sigma_z$ vs $\varepsilon_x$ (b), and  $\sigma_z$ vs $\sigma_x$ (c).  Cases with 5k, 500, 200, and 100 training points are shown from top to bottom.  500 randomly-sampled training points are sufficient in this case for obtaining an accurate estimated Pareto front with the NN model. For 200 training points, the estimated Pareto front is quite a bit less accurate but still could provide an initial population for subsequent fine-tuning with a GA run on the physics simulation.\label{fig:num_needed2}} 
\end{figure}

To address this, we trained models using 5000, 500, 200, and 100 randomly-sampled points for the \SI{40}{nC} case and compared the resulting estimated Pareto fronts (see \figref{fig:num_needed2}). Note that when using only 500 points, we do not see substantial reduction in the estimated Pareto front accuracy, compared with the 5000 point case. For training with 200 points, the prediction starts to deviate substantially from the ground truth, but the solution could still be used used as an initial guess (or 'warm start') for subsequent fine-tuning with the physics simulation (i.e. by including new points from the estimated front in the training set and iteratively retraining).

\begin{figure}[h!]
	\centering
	{\includegraphics[width=0.3\textwidth]{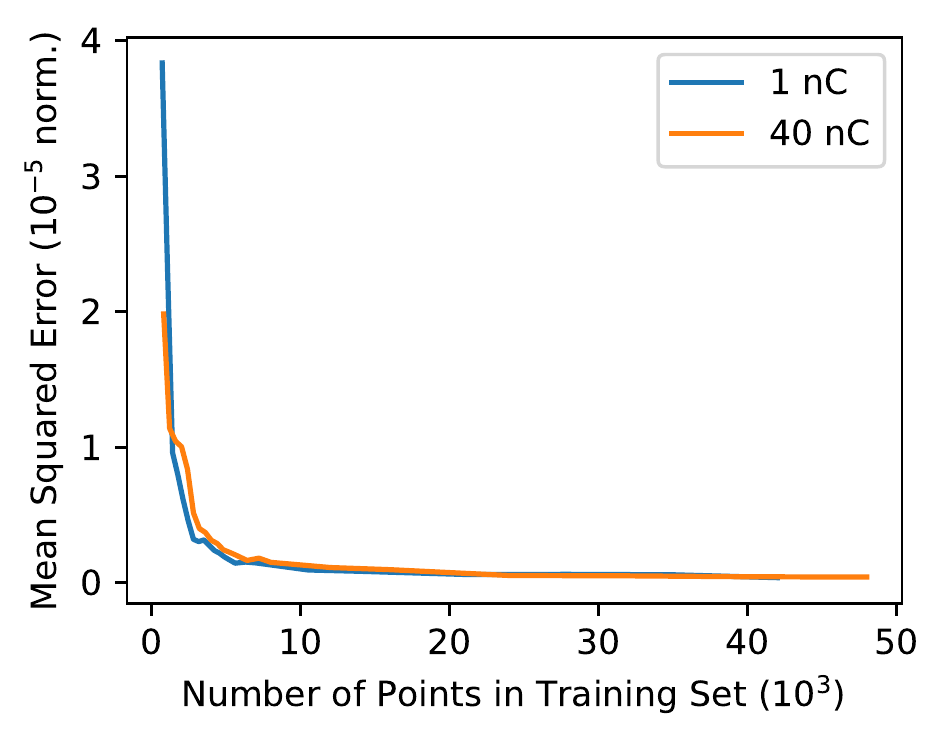}}

	\caption{Prediction error in normalized units for AWA model trained on various training set sizes. After a few thousand points, we see diminishing returns. \label{fig: trainsize}}
\end{figure}

We also trained the NN models on a larger range of training sample sizes and evaluated their performance in predicting the points obtained from running the GA with the physics simulation. We quickly see diminishing returns in improvement for the prediction task after the number of samples increases beyond a few thousand. In \figref{fig: trainsize} we show the impact of changing the training set size on the prediction performance.

\begin{figure}[h!]
	\centering
	{\includegraphics[width=0.7\textwidth]{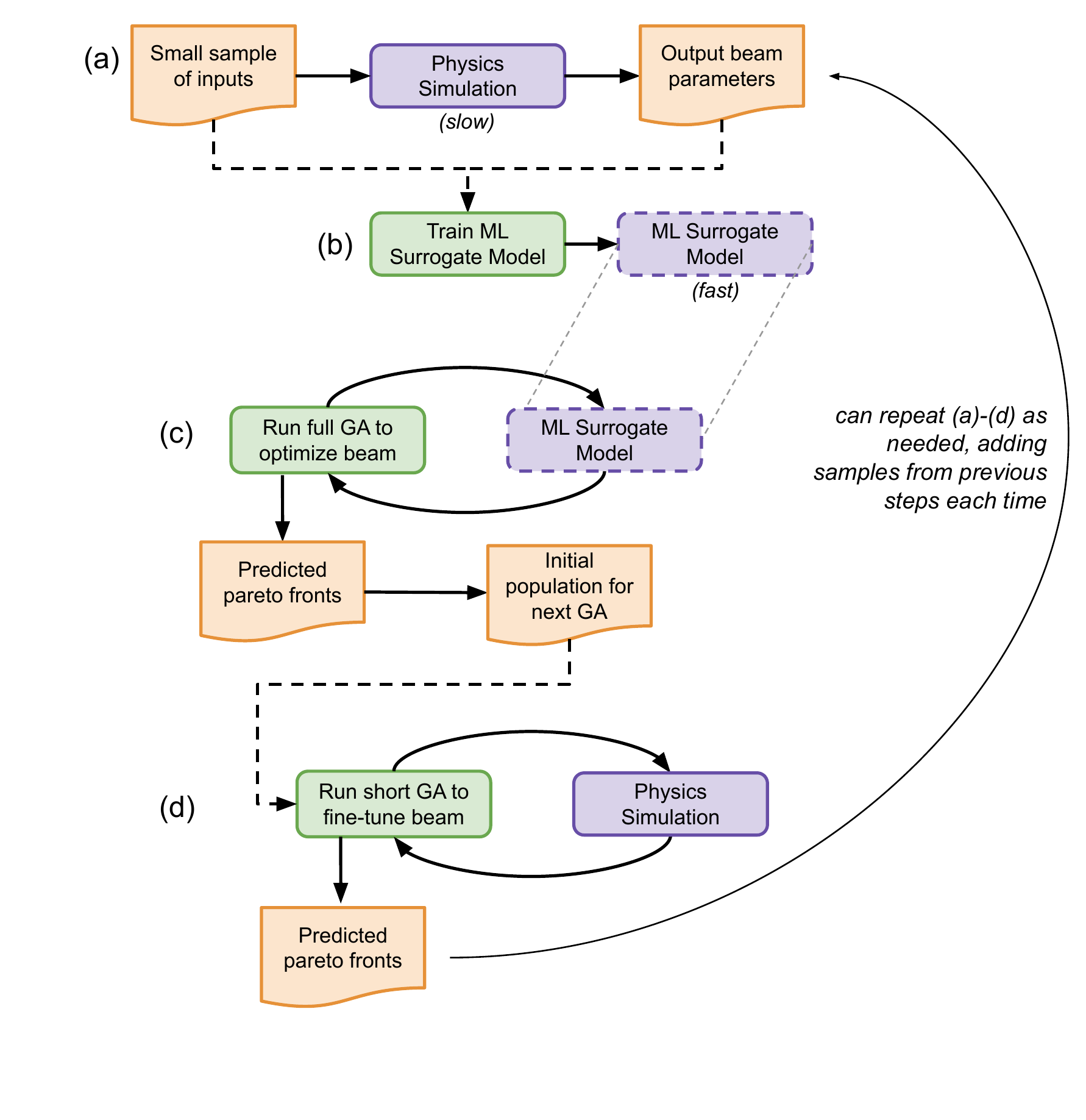}}
	\caption{Workflow for a method to obtain orders of magnitude faster \moo\ optimization of particle accelerators. First, run a small random sample of inputs through the \psim\ (e.g., a few hundred points in our case) (a). Train an ML model on the random sample (b). Run the \moo\ with the ML model to obtain predictions of the estimated Pareto front and corresponding optimal input settings on the machine (c). Use the predicted optimal input settings as the initial population in a second, shorter \moo\ optimization over the \psim\ to fine-tune the result (d). One can also incorporate the new data into a second training of the ML model and repeat these steps as needed until convergence (i.e. iterative retraining and optimization). \label{fig: new_meth}}
\end{figure}

Note that the uniform random sampling of the input space does not map onto the hypervolume of the output space evenly, and instead we see clusters of points in the output space (for example, see the $\Delta$$E$ vs. $\varepsilon_x$ plots in \figref{fig: example-par}). The consequence is that having just a few points for some large parts of the output space results in a model that does not represent these regions as well. It also suggests that the space could be sampled more efficiently (e.g. to avoid over-sampling some regions and under-sampling others). This points to the potential utility of a more intelligent sampling strategy. To see whether iterative retraining might be a viable approach to reduce the number of required samples, we conducted a preliminary study. For this we take a small initial sample, train the model, run a GA on the NN model, evaluate a small set of the resultant solutions, add these to the training set, and then repeat these steps until convergence. This process is shown in \figref{fig: new_meth}. 

\begin{figure}[h!]
	\centering
	{\includegraphics[width=0.35\textwidth]{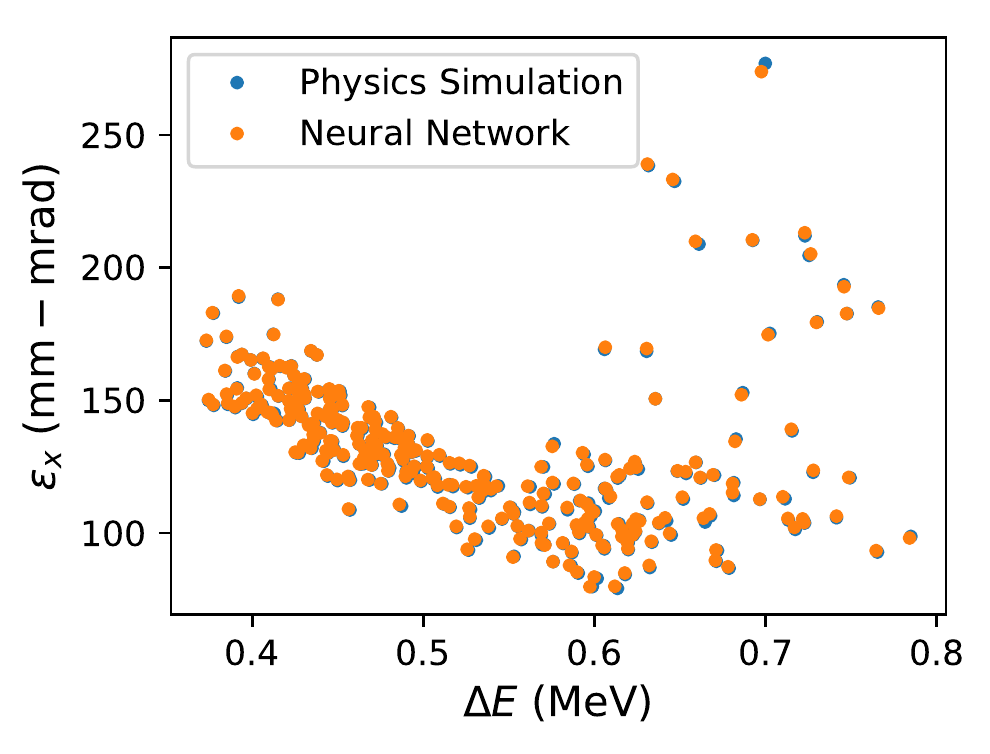}
	\includegraphics[width=0.35\textwidth]{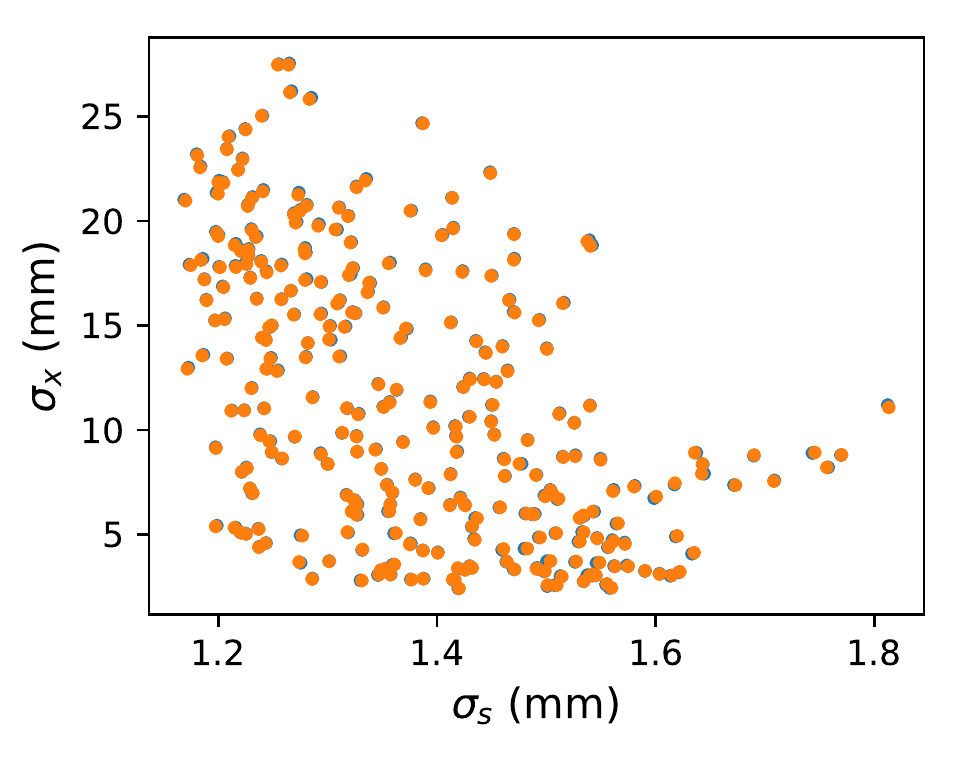}
	\includegraphics[width=0.35\textwidth]{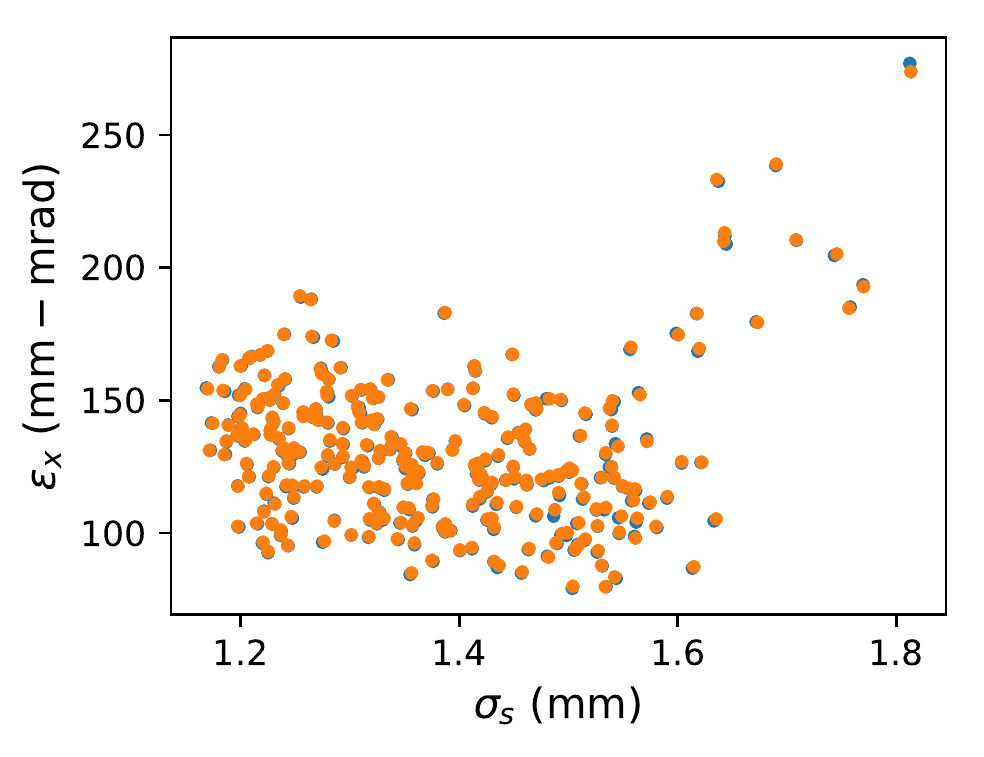}
	}
	\caption{Point predictions from the NN surrogate model and the corresponding points from the physics simulation, showing that the two are in excellent agreement. These samples are drawn from the test set. \label{fig:surro_ga}} 
\end{figure}

To explore the iterative retraining approach, we leveraged the fast execution of the NN model that was trained on all the random sample data (referred to as the `NN surrogate' in the following text) to provide a proxy for the true function. We did this because it would be too computationally intensive for us to explore this approach with the physics simulation directly, particularly when taking into account the number of trials that we anticipated would be needed to refine the method. It also highlights an interesting use of the NN surrogate model: prototyping optimization algorithms. To ensure that it is acceptable to use the NN surrogate as the proxy function in this case, we compared point predictions from it with those from the physics simulation and find that they are in excellent agreement (for example, see \figref{fig:surro_ga}). The convergence of the GA run on the NN surrogate also matches the convergence of the GA run on the physics simulation. Beyond this, we also trained a model on actual input and output data from the physics simulation and another model on output data generated by the NN surrogate. In this case we used a 350-point sample so that we were below the threshold that was needed to obtain an accurate front. We then compared the convergence of the GA on each of these models and found that they were the same. Taken together, this provides assurance that we can use the NN surrogate as a proxy for the physics simulation in the iterative retraining study.

For the study, we conducted iterative updates to a second, randomly-initialized NN model and observed how many function evaluations were needed to obtain an estimated Pareto front that matches the one obtained from the physics simulation. Note that here we are not directly picking points on the front to add to retraining (which could also be done), but rather, we are selecting a set of additional points using the NSGA-II selection algorithm (i.e. the best-performing solutions with respect to all 7 output objectives). We start with a random sample of 20 points, train the NN model, run a GA with 500 generations and 100 individuals in each generation, select 20 points from the final population, run these points through the full NN surrogate (which provides an accurate proxy for the physics simulation in this case), add these new evaluated points the training and validation sets (16 to the former and 4 to the latter), and then repeat the steps. We repeated this process with 10 random sets of initial points and found that for \SI{40}{nC} we were typically able to reduce the number of required points to obtain an accurate front to between 120 and 200 in total (i.e. 6--10  iterations of the algorithm). The variance is due to differences in where the 20 initial randomly-sampled points are located (i.e. some of these random initializations will happen to map the output space more evenly than others). The number of generations and individuals to use for the GA were selected by doing a small search of the GA hyperparameters. Investigating this in more detail (e.g. rigorously assessing the impact of hyperparameters for both the GA optimization and the NN training) is outside the scope of this paper. However, these results affirm the intuition that iterative retraining is a viable option for substantially reducing the number of required training points.


\begin{figure}[ht!]
	\centering
	{\includegraphics[width=0.25\textwidth]{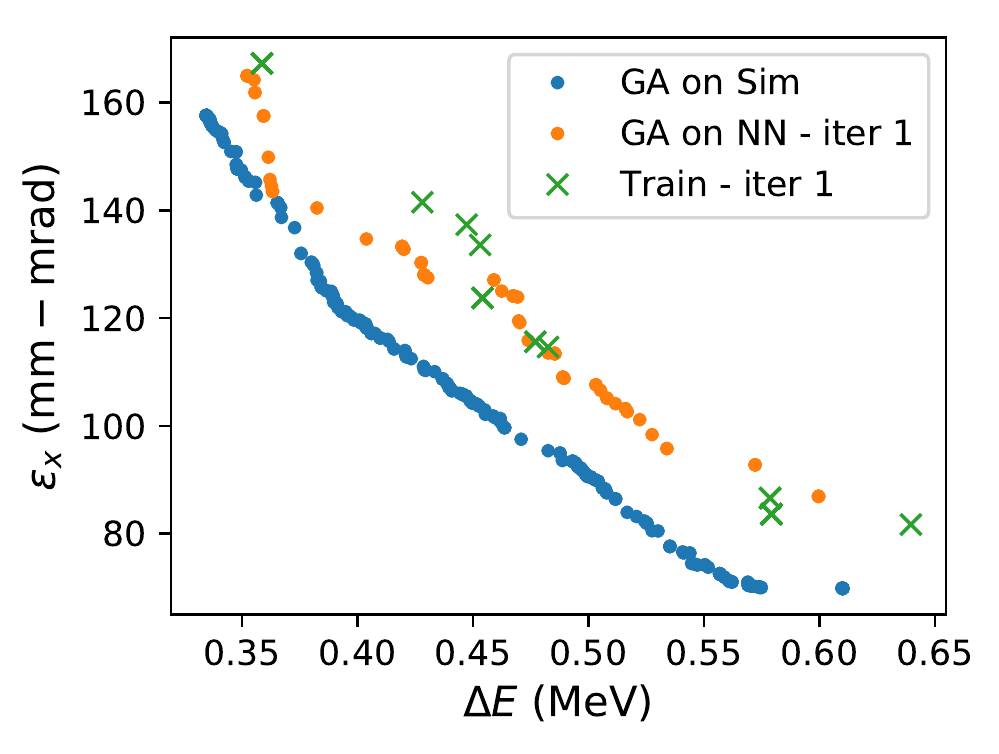}
	\includegraphics[width=0.25\textwidth]{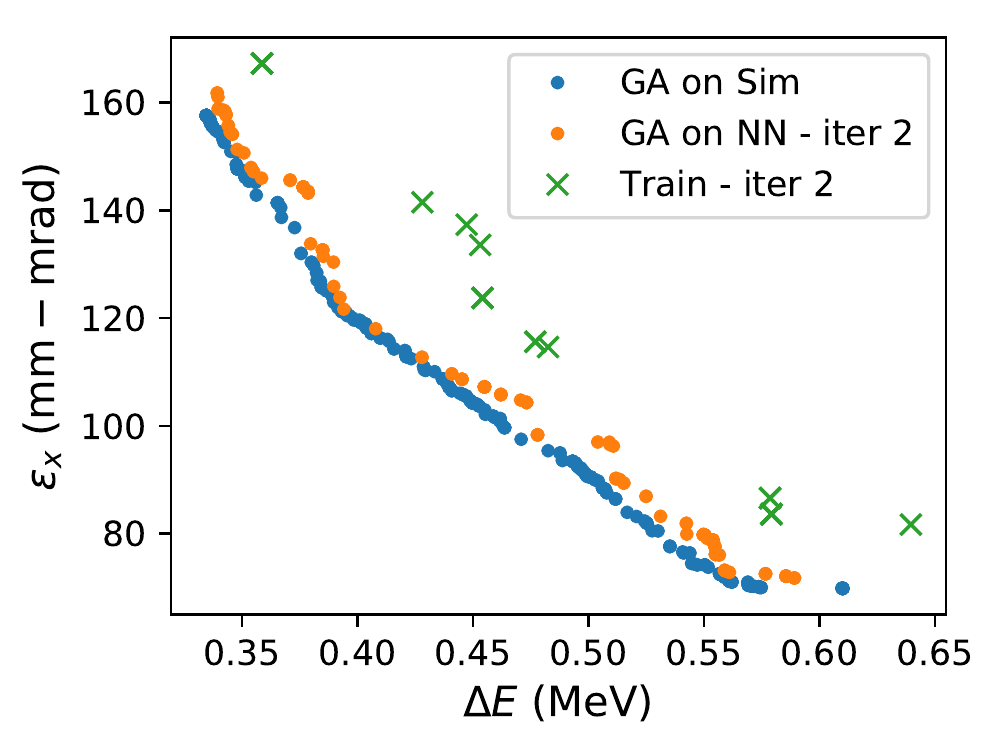}
	\includegraphics[width=0.25\textwidth]{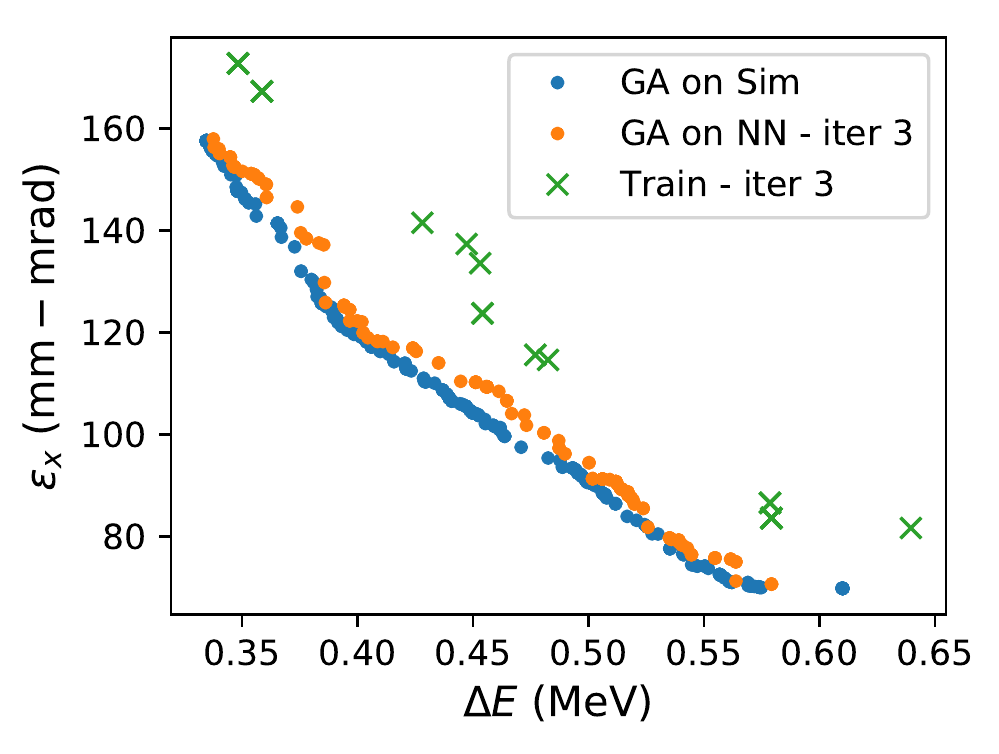}
	\includegraphics[width=0.25\textwidth]{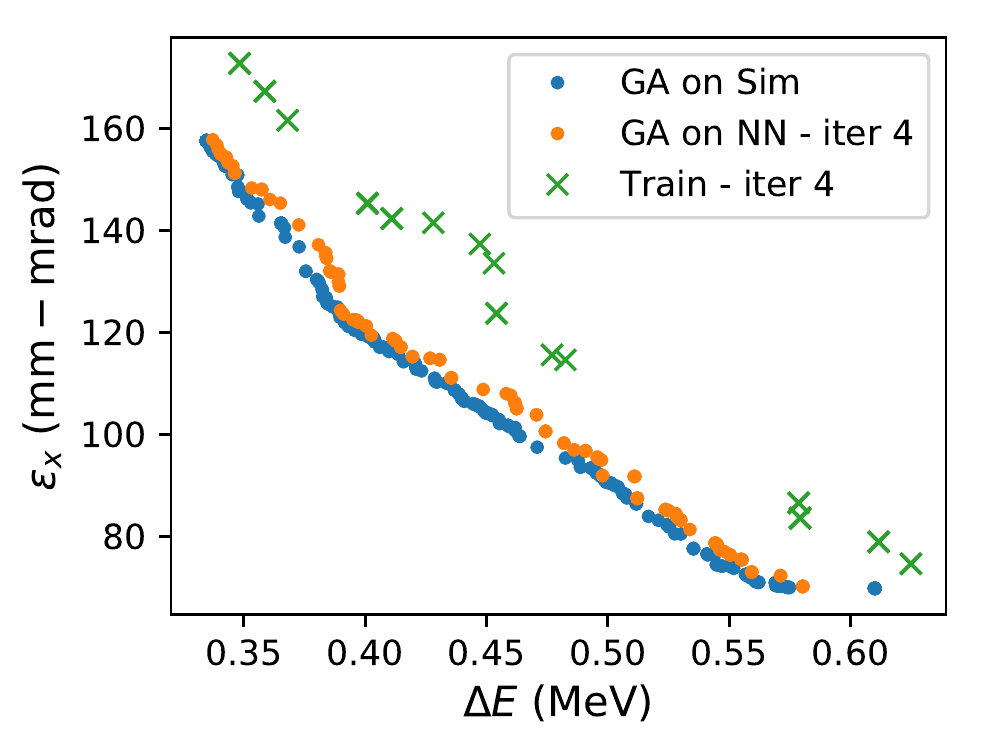}
	\includegraphics[width=0.25\textwidth]{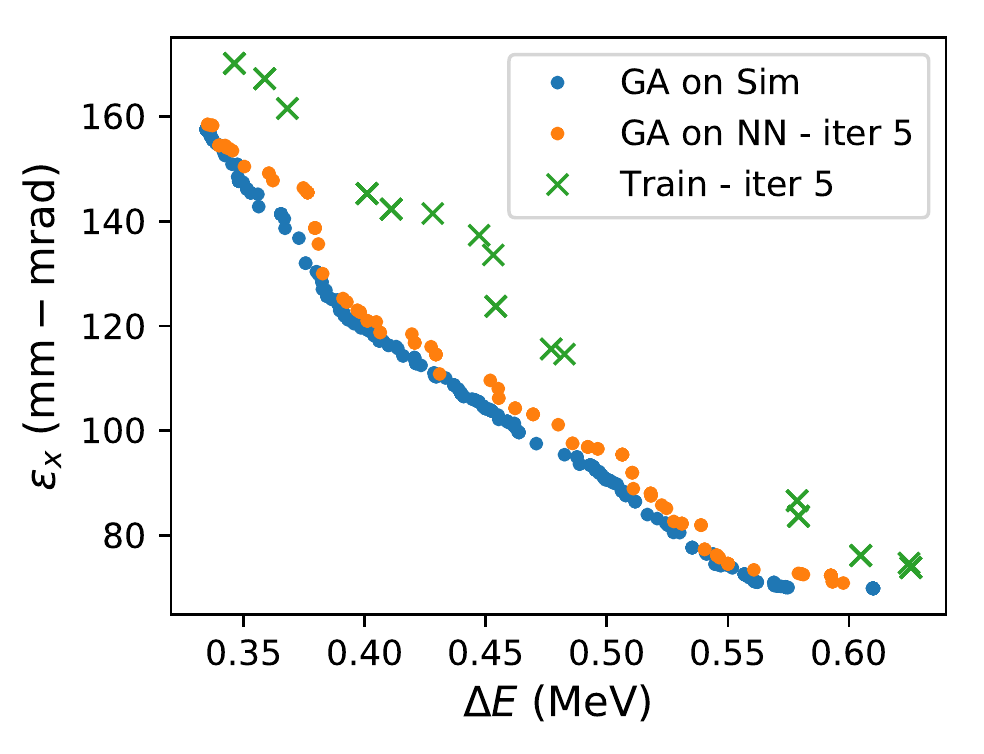}
	\includegraphics[width=0.25\textwidth]{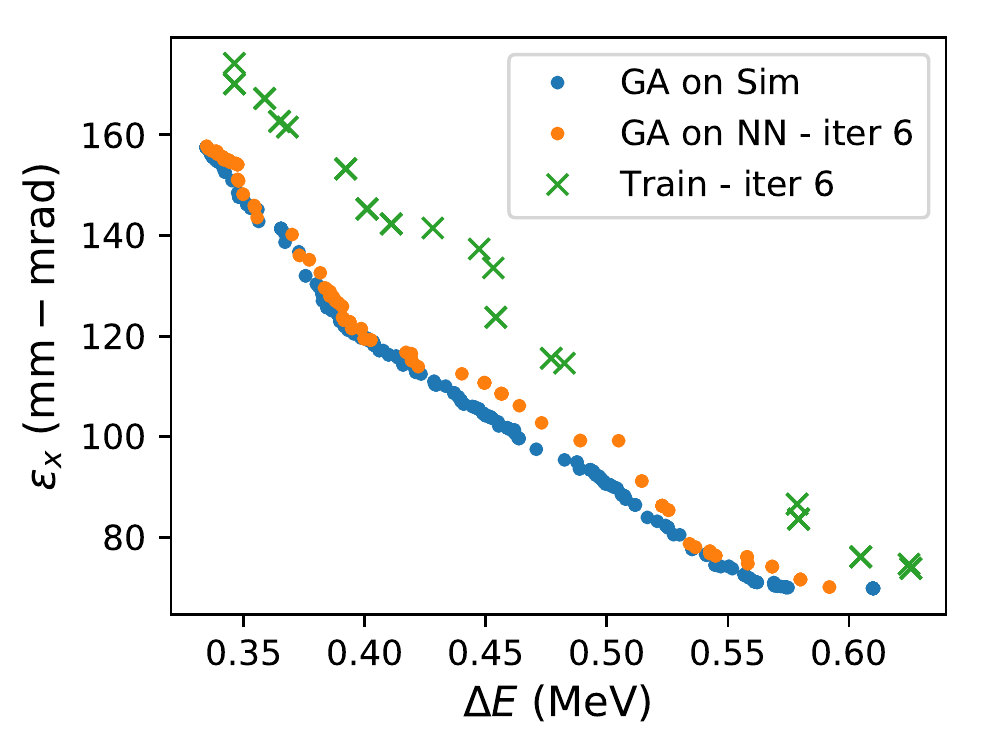}
	\includegraphics[width=0.25\textwidth]{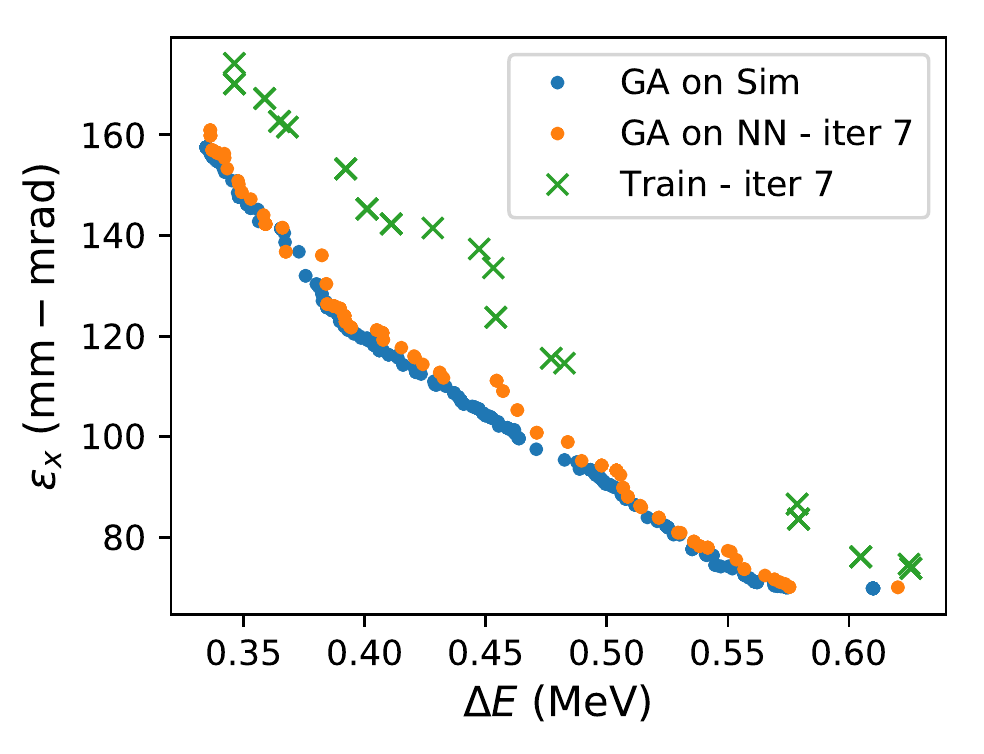}
	\includegraphics[width=0.25\textwidth]{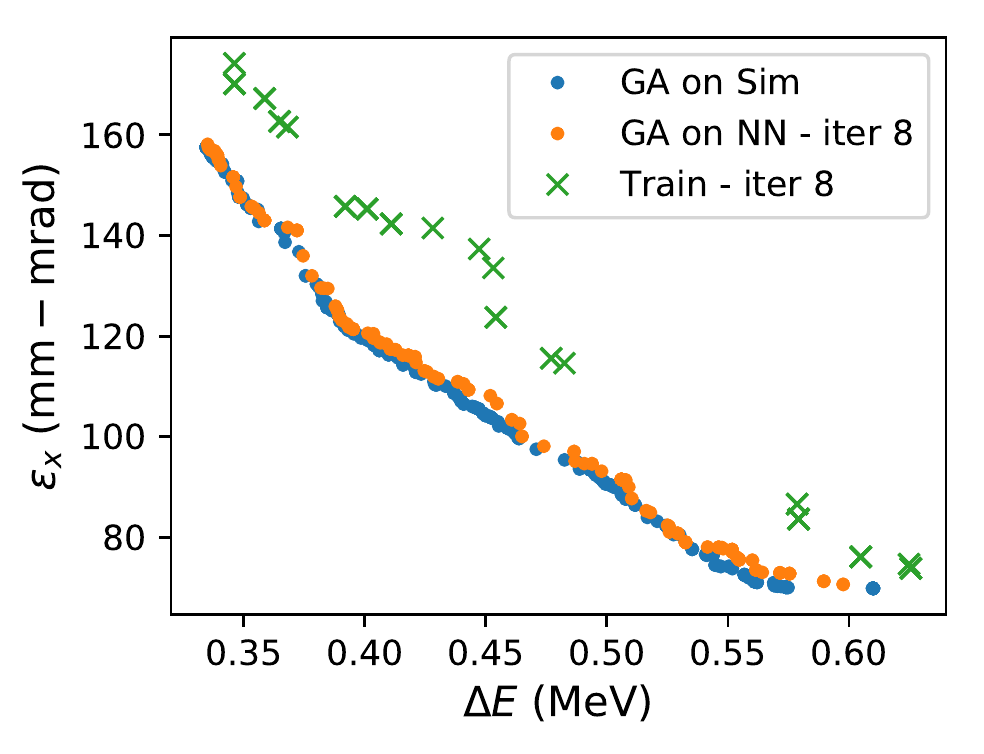}
	\includegraphics[width=0.25\textwidth]{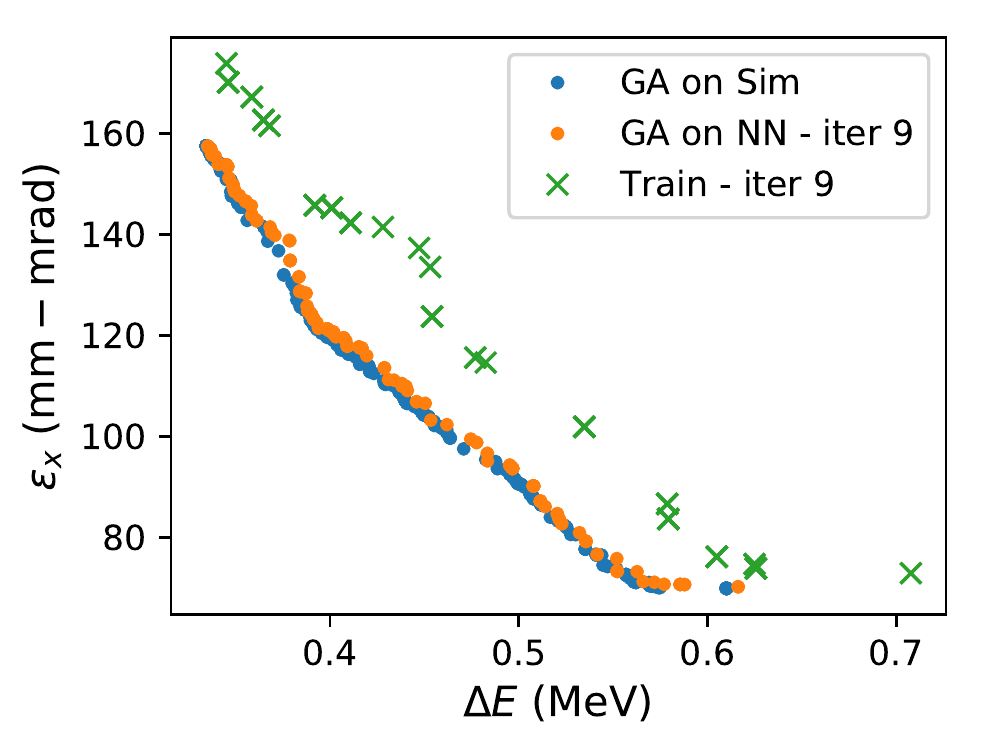}
	\includegraphics[width=0.25\textwidth]{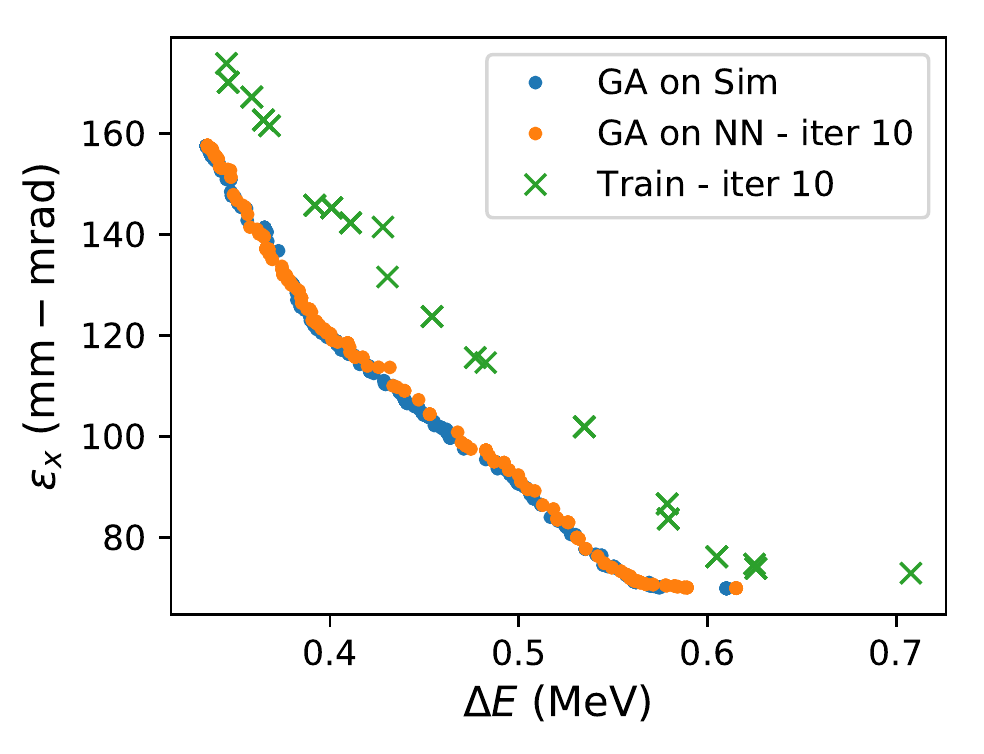}}
	\caption{Example of convergence of the estimated front  for $\Delta$$E$ vs. $\varepsilon_x$ during iterative retraining. Here we show 10 iterations from one of the trials and display the estimated Pareto front projections from the physics simulation, from the NN model on each iteration of the retraining process, and from the training data used in that iteration. Substantially fewer points (e.g. 200 in this run) are needed to obtain an estimated front that matches the one obtained with the physics simulation, as compared with using a single random sample for training. Note that this example is one of the slower-to-converge runs, which we deliberately chose as a conservative demonstration of the method (i.e. other runs had a fewer number of required points, but that relies on lucky placement of the initial random sample points). \label{fig:iter_retrain}} 
\end{figure}

\subsection {Improvement in Computational Efficiency by Using ML Model}
Once training is completed, one NN model evaluation can be computed in $<$ 1 ms on one core of a \textit{laptop computer}, compared with 590 seconds for one \psim\ on 8 cores for the \SI{40}{nC} simulation. In \tabref{tab: speedup1}, the time-to-solution and computing resources needed for the GA optimization with the \psim\ are compared with those needed for GA optimization with the NN surrogate model. The NN-based \moo\ takes about 2 minutes on a laptop computer, which corresponds to $\mathcal{O}(10^{6})$ times fewer computing resources (in terms of core-hours) than the same optimization when conducted with the \psim\ directly. 

Importantly, the overall improvement is still substantial when considering the computation time required to generate the training data and to train the NN. This makes the approach a viable way of speeding up initial design optimization. To generate enough training data to produce an accurate estimated Pareto front in the \SI{40}{nC} case, 132 times fewer simulation evaluations and 144 times fewer total core hours were required than if one were to use the \psim\ alone. Finally, the NN training itself takes approximately 10 minutes on one core of a laptop. Furthermore, for a given problem, this step of generating the data and training the model only needs to be done once, and the NN model can be used for subsequent modeling and optimization tasks (including serving as a fast stand-in for part of a larger start-to-end simulation of the full machine). 

Note that although we showed results in the previous section that suggest there would be further reduction of the computational resources by conducting iterative retraining, we do not include results for this in \tabref{tab: speedup1} because we did not repeat this with the physics simulation itself. On average, the cost of running each iteration of iterative retraining included 33.5 seconds for training and 43.0 seconds for running the GA (with standard deviations of 34.7 and 24.9 respectively). For a case with 10 iterations of convergence, this would add 0.21 core-hours on top of the cost of running the simulation and allow a lower number of simulation samples to be used (likely about 264 core-hours of simulation time instead of 660). This corresponds to a factor of 360 times fewer core-hours than were needed to obtain the equivalent solution with the physics simulation alone. The  NN surrogate is an accurate proxy for the physics simulation (and thus this result is encouraging), but the actual computation time should be verified in future work by running the same iterative retraining procedure with the physics simulation directly.

\begin{table}[tb]
\caption{Comparison of computing resources: core-hours, wall time, and number of simulation evaluation required for running the \moo\ with the \psim\ and the \moo\ with the NN. Here we show only the \SI{40}{nC} bunch charge, but details on \SI{1}{nC} bunch charge can be found in the Appendix \label{tab: speedup1}. When running the same GA with the NN, $3 \cdot 10^{6}$ times fewer computing resources are required. When including the resources needed to generate the training data (as might be done for initial design optimization), we still have a factor improvement of 144 in terms of core-hours required. Note that although we showed results in the previous section that suggest there would be further reduction of the computational resources by conducting iterative retraining (e.g. 360 times fewer core-hours for 10 iterations), we do not include results for this in the table because we did not repeat this with the physics simulation itself.}
\begin{tabular*}{\columnwidth}{@{\extracolsep{\fill}}llccc}
\toprule
\textbf{Method}		& \textbf{Calculation } 	& \textbf{Core-hours}  & \textbf{Wall time (hours)}  & \textbf{Evaluations}    \\
\midrule
Physics Sim. 	& GA on OPAL					 &	95,000			& 36 	& 65,929 \\
\midrule
ML-based 	&Generate training data 	 &	660			& 0.33		& 500 \\
			&Train NN                         & 0.17      & 0.17  & \textit{n.a.}  \\
			&GA on NN					&	0.03			& 0.03	& 65,600 \\  
			&\textit{Speedup - training included}	&	\textit{144$\times$} 			& \textit{109$\times$}	& \textit{132$\times$} \\  
			&\textit{Speedup - training excluded}	&	\textit{3$\cdot10^{6}$$\times$} 	& \textit{1200$\times$}	& \textit{n.a.} \\  
\bottomrule
\end{tabular*}
\end{table}

When considering the reduction in computational resources used, it is important to note that the \psim\ had already been tuned for a high level of computational efficiency and we ran the simulations in parallel. As such, the results represent improvement over the state-of-the-art.  In cases where simulation codes are less efficient or higher complexity (e.g. plasma simulations, FEL simulations), the ML-based approach may actually enable larger optimization studies to be conducted than would have been feasible using the \psim\ alone.

\clearpage
\subsection {Comparison with Different ML Models}
As a check to see whether a linear model would be sufficient for this problem, we trained a support vector regression (SVR) model~\cite{vapnik1998statistical} with a linear kernel. For the error metric, we used the mean squared error (MSE), defined as MSE = $\frac{1}{N}\sum_{i=1}^{N}(y_i - \hat{y}_i)^2$, where ${N}$ is the number of samples, ${y_i}$ is the true output for sample ${i}$, and ${\hat{y}_i}$  is the predicted output for sample ${i}$.  The MSE over all predicted beam parameters using the SVR model was $5.5\times10^{-6}$, in comparison to $3.5\times10^{-10}$ for the NN model, indicating that we do gain accuracy by using a nonlinear model. Note that although these differences in error may at first strike readers as small, we are dealing with quantities like emittances that have raw values on the order of $10^{-6}$ m-rad and rms beam sizes that have raw values on the order of $10^{-3}$ m. For example, for an error in emittance of $0.1\times10^{-6}$ m-rad we would expect a squared error on the order of $10^{-14}$. Considering this, the difference between the MSE of the SVR and NN model is substantial.

\begin{figure}[ht!]
	\centering
  	{\includegraphics[width=0.9\textwidth]{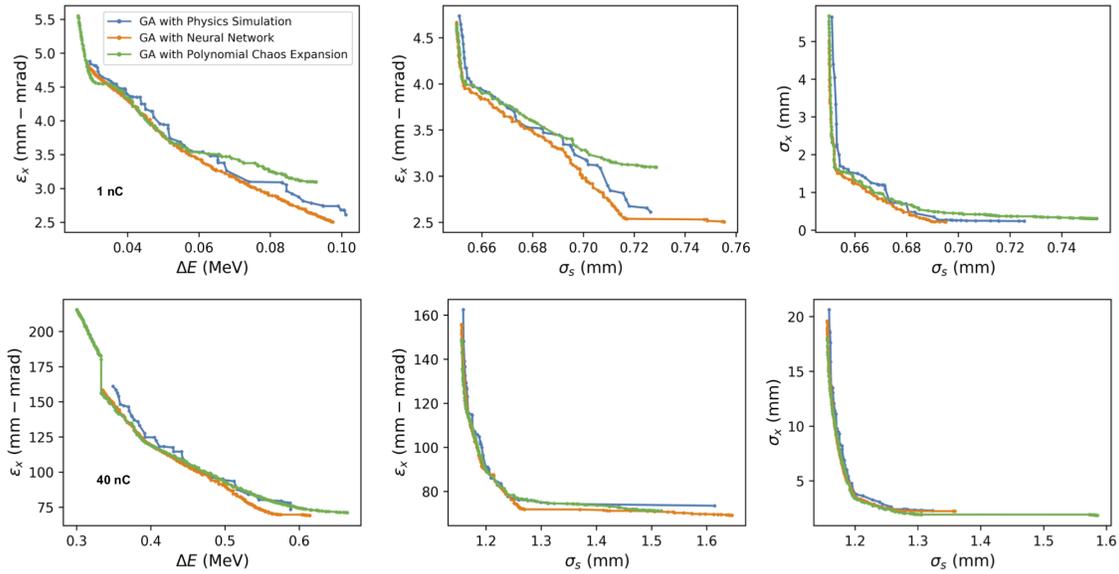}}
  	\caption{Comparison of the estimated Pareto front for a PCE model and a NN model. Note that in this case, we use a random sample of 42,000 points (this was from the large initial random sample, which we ran before attempting to scale down to as few sampled points as possible during training). To make a fair comparison in this case, we use the same number of sample points for the ANN as we use for the PCE. Note that although the PCE model does not perform as well, it is more straightforward to train, has fewer hyperparameters to tune, and can be used to provide an uncertainty estimate. Although the fronts are less accurate, the points could be used as an initial starting point for a subsequent GA run on the physics simulation. \label{fig: pareto_compare_all}}
\end{figure}

We also compared performance of NN and PCE surrogate models, both in parameter prediction and in the optimization task. In \figref{fig: pareto_compare_all}, we show a comparison between the estimated Pareto fronts obtained with the PCE and the NN models. The MSE of the PCE model for \SI{40}{nC} was $1.6\times10^{-7}$ and for \SI{1}{nC} was $4.5\times10^{-6}$. In contrast, the NN models were more accurate with a MSE of $3.5\times10^{-10}$  and $2.9\times10^{-8}$ on the  \SI{40}{nC} and \SI{1}{nC} cases, respectively. However, one downside of using a simple NN model is that it does not inherently give an estimate of prediction uncertainty and model sensitivity without additional analysis. In contrast, the PCE model has the benefit of providing straightforward estimates of prediction uncertainty and sensitivity via the \sobol\ indices~\cite{Sobol01,aa1}. The PCE model also has fewer hyperparameters to tune (i.e.,\ polynomial order, type of polynomial used). Although the estimated fronts are less accurate, the points from the PCE could be used as an initial starting point for a subsequent GA run on the physics simulation.


\clearpage
\section* {Extension to a Challenging Cyclotron Example} 

\begin{figure}[ht!]
	\centering 
  	{\includegraphics[width=0.6\textwidth]{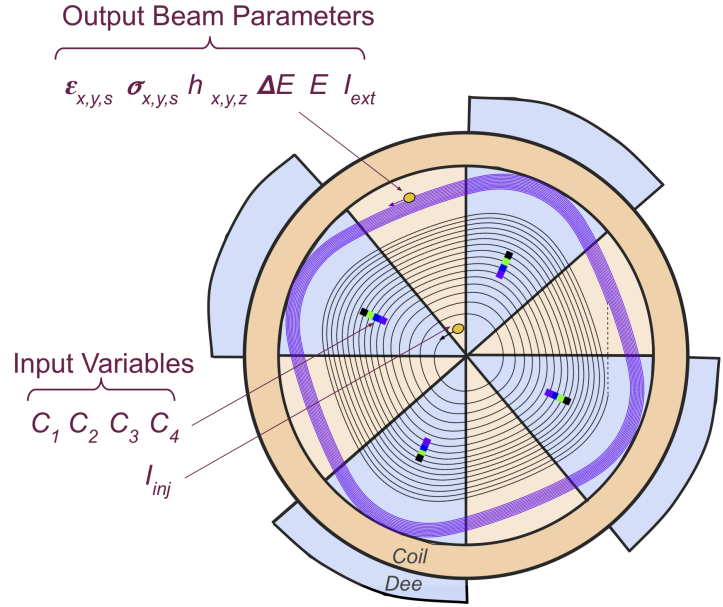}}
  	\caption{The IsoDAR schematic, showing an overlay of the magnetic field configuration (coil \& dee) and central particle trajectories from the simulation. The radio frequency system used for acceleration is not shown. In
	black, the centroid trajectories of 18 turns are shown. A particle bunch of intensity $I_{\text{inj}}$ is injected at the centre and passes through 4 families of collimators (indicated with the colored rectangles starting at turn 6). Each collimator family has
	one parameter $C_{1} \ldots C_{4}$ defining the collimator aperture. The collimator aperture and the initial intensity
	are input variables used by the optimizer. In this schematic, the extraction system is not shown, nor are all $100$ turns of the beam. The last 10 turns before extraction are indicated in purple, together with the output beam parameters: the halo parameters $h_{x,y,z}$, the beam size $\sigma_{x,y,s}$, the projected emittance  $\epsilon_{x,y,s}$ and the extracted beam current $I_{\text{ext}}$. Particle losses ($I_{\text{inj}}-I_{\text{ext}}$) in the collimators have to be minimized, but at the same time a small $h_{x}$ is desired. This is one of the most
important considerations in the design phase.  \label{fig: isodar_diagram}} 
\end{figure}

To assess this approach further, we applied it to a higher-dimensional and generally more complex accelerator problem: the IsoDAR cyclotron, shown in  \figref{fig: isodar_diagram}. The IsoDAR cyclotron will provide ~60 MeV protons for sterile neutrino search \cite{PhysRevLett.109.141802}. It has also been proposed to use this machine to produce medical isotopes with high efficiency \cite{isodarnature}. In high-power proton machines, a major challenge is to minimize the number of uncontrolled, lost particles (thus maximizing the number that are sent to the relevant experiment and also minimizing damage to the machine). An indication for losses are large halos around the core of the particle beam. The halo can be quantified by a set of halo parameters and is proportional to the kurtosis, which is a measure of the ``tailedness" of the probability distribution. A detailed description of this problem can be found in \cite{aa1}.
 
The IsoDAR cyclotron simulation is implemented in OPAL and includes nonlinear beam collective effects such as the beam self-fields. This simulation is an order of magnitude more expensive to evaluate than the AWA due to the long path the particles travel and correspondingly long integration time. We also attempt to predict a more difficult set of beam parameters than we did for the AWA. The most difficult parameter to predict is the beam halo, which is notoriously challenging to simulate accurately. Because the simulation is so computationally intensive, we use only a small number $N$ of particles (i.e.,\ $N=1.33\times 10^{5}$). As a result, we expect that the actual prediction of the halo is
not fully converged (which should show up as noisy estimates of the halo).

Using a 2500-point random sample, we trained a NN to predict 12 beam parameter outputs: the total number of lost particles ($P_L$), the beam energy ($E$), energy spread ($\Delta$$E$), transverse and longitudinal beam sizes ($\sigma_x$,  $\sigma_y$, $\sigma_s$), transverse and longitudinal emittances ($\varepsilon_x$, $\varepsilon_y$, $\varepsilon_s$), and the beam halo ($h_x$, $h_y$, $h_z$). Again, definitions of the beam parameters can be found in \tabref{tab: beamparam} in the Appendix.  Here we use the definition of the halo parameter found in \cite{PhysRevSTAB.5.124202}, formula 12. The 5 input parameters are the initial proton beam current and the positions of four families of collimators (corresponding to 16 collimators in total). For the GA run on the NN, we use 1000 generations with 300 individuals. Other than this, the implementation is the same as that used for the AWA. 

\begin{figure}[ht!]
	\centering
  	{\includegraphics[width=0.9\textwidth]{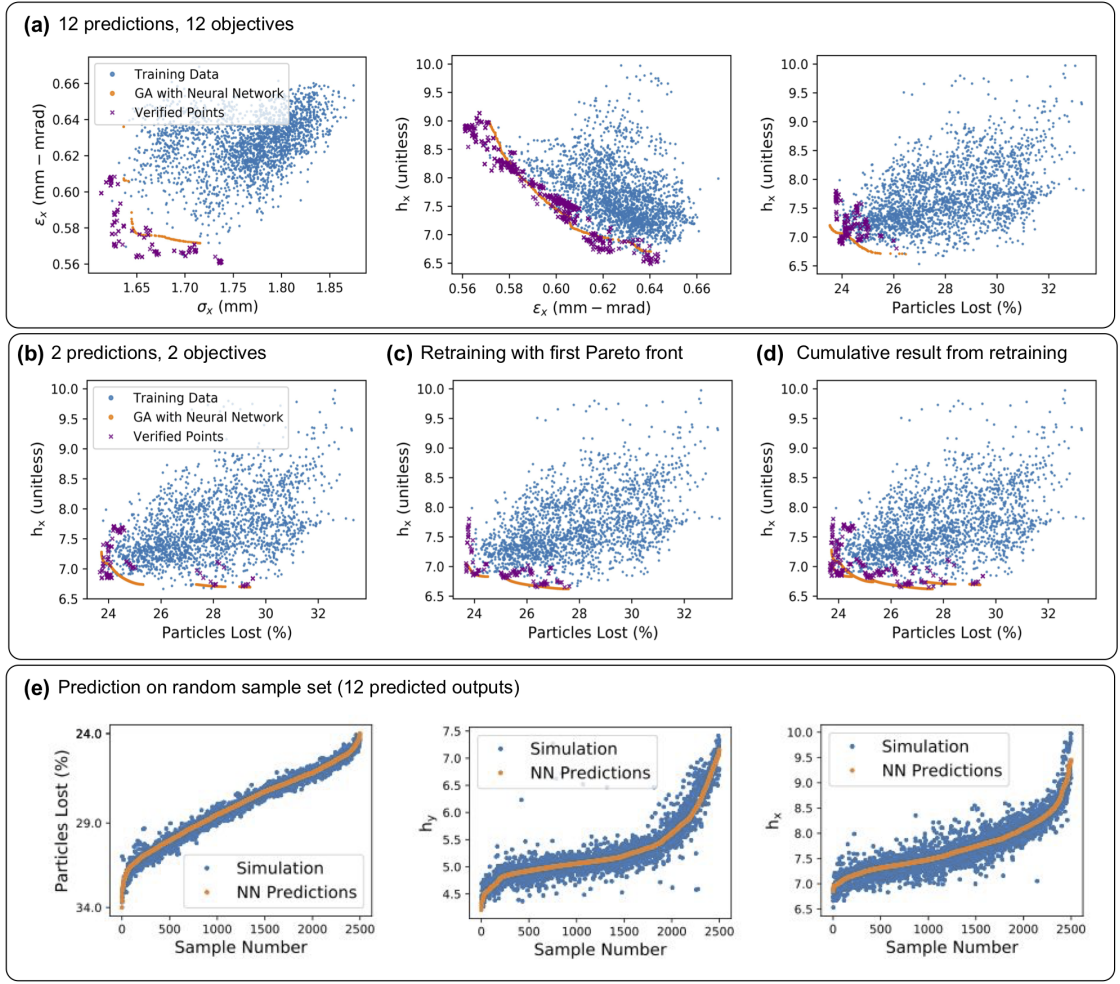}}
	
  	\caption{Example of estimated Pareto fronts from IsoDAR NN model and corresponding verified points from the physics simulation. We first show results from training a model to predict and optimize 12 beam parameter outputs (a). We then show corresponding results from training only on the two most challenging beam parameter outputs,  $h_x$ and $P_L$ (b). We then include the verified front from the previous model in the training set, retrain the model, repeat the optimization, and check the new Pareto points (iterative retraining) (c).  Although the fronts do not match as well as they do for the AWA, for this challenging problem it is encouraging that we can predict roughly accurate fronts. Finally, we show the prediction on sorted values of the random sample set. \label{fig: pareto_isodar}} 
\end{figure}

As expected, upon verification with the simulation we find that the estimated Pareto fronts (see \figref{fig: pareto_isodar}) are not as accurate as those obtained for the AWA. However, the points still roughly map out the region of parameter space near the estimated Pareto front from the GA run on the physics simulation. It is unclear how much of this is a contribution from the simulation itself, since it has not been tuned to ensure that the mesh size and number of particles is sufficient to produce a reliable result. Fine-tuning of the simulation was not conducted because the associated computational intensity of conducting this process for the broad range of parameters we are varying was too high to be feasible given the compute time available to us. Note that in this case, we do not have an estimated Pareto front from running the GA on the physics simulation. This is because the IsoDAR simulation is so computationally expensive that we were unable to run a multiobjective optimization on all parameters to convergence. All this taken into account, we consider finding verified points that are in close proximity to the Pareto front estimated by the NN to be a successful use of the approach.

For the tradeoff between halo parameters and losses, we obtain results that more closely outline the extent of the observed hypervolume projection when we produce a model that only predicts those direct outputs (as opposed to predicting and then optimizing all 12 objectives). In \figref{fig: pareto_isodar}, we show this for $h_x$ and $P_L$. This is also more in line with how GAs are typically used by accelerator physicists at present: often, only two important parameters will be optimized as competing objectives. We also find that by iteratively retraining the model with the verified points and repeating the process with the new model, we obtain a model that fills out more of the front (also shown in \figref{fig: pareto_isodar}).

\begin{figure}[ht!]
	\centering
	{\includegraphics[width=0.3\textwidth]{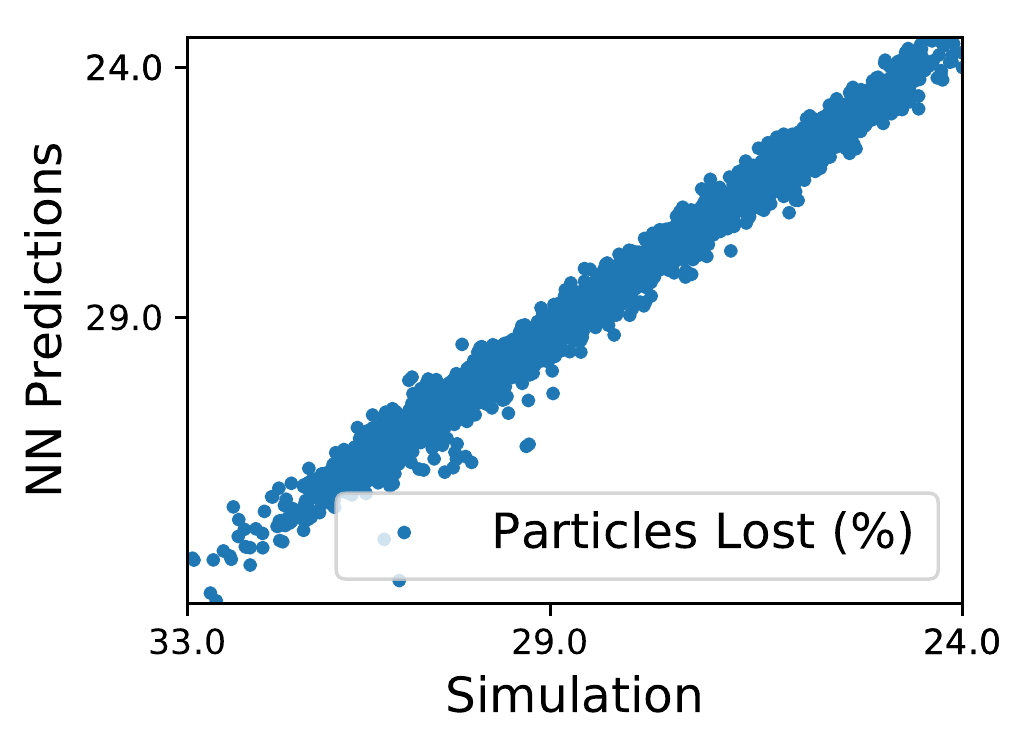}
	\includegraphics[width=0.3\textwidth]{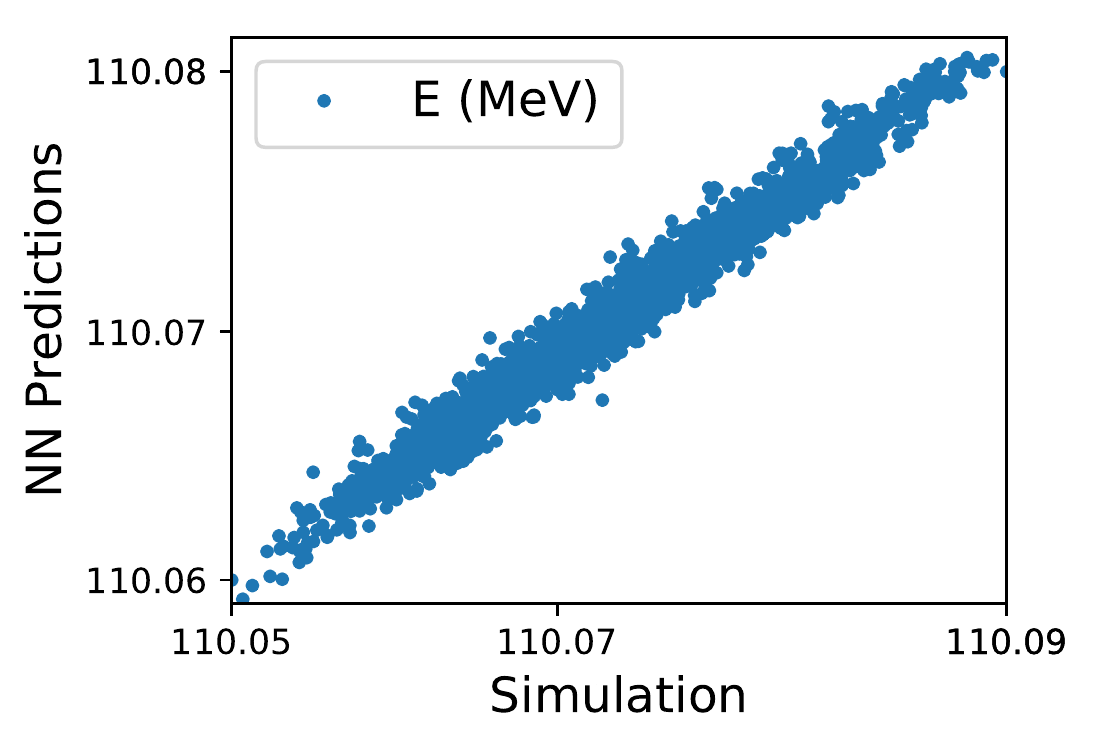}
	\includegraphics[width=0.3\textwidth]{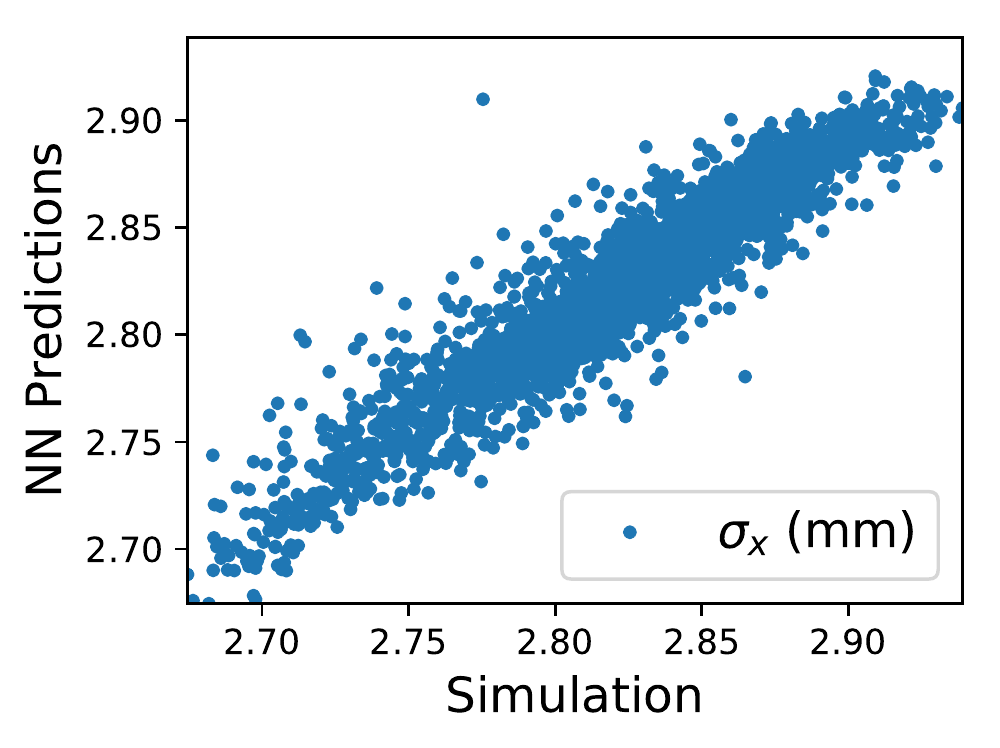}
	\includegraphics[width=0.3\textwidth]{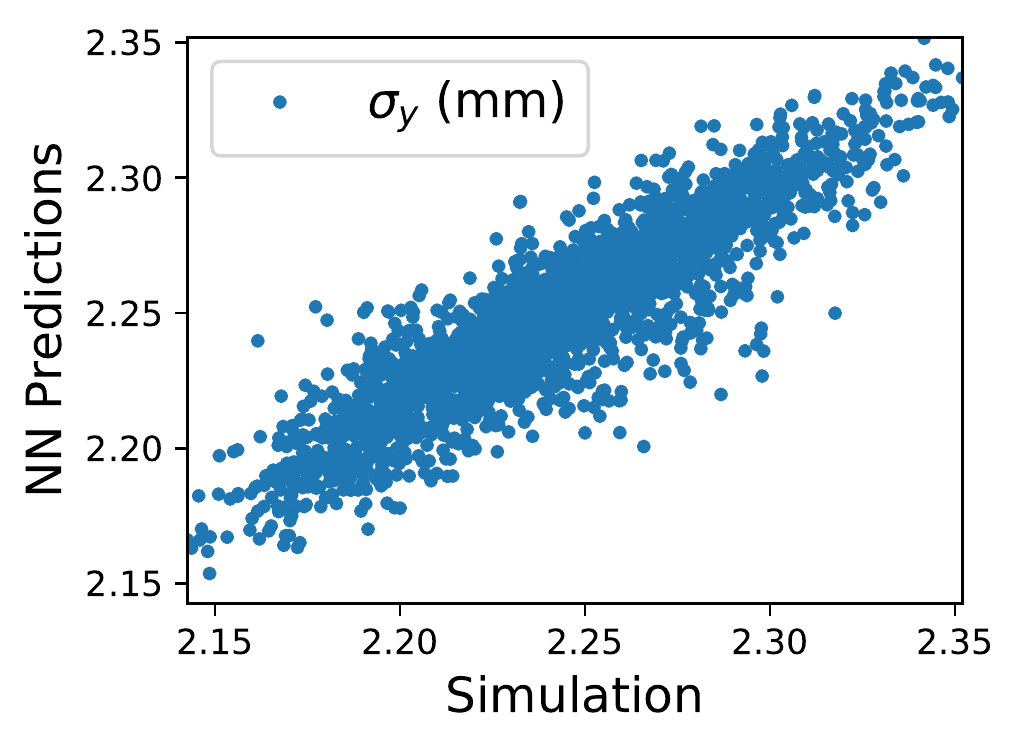}
	\includegraphics[width=0.3\textwidth]{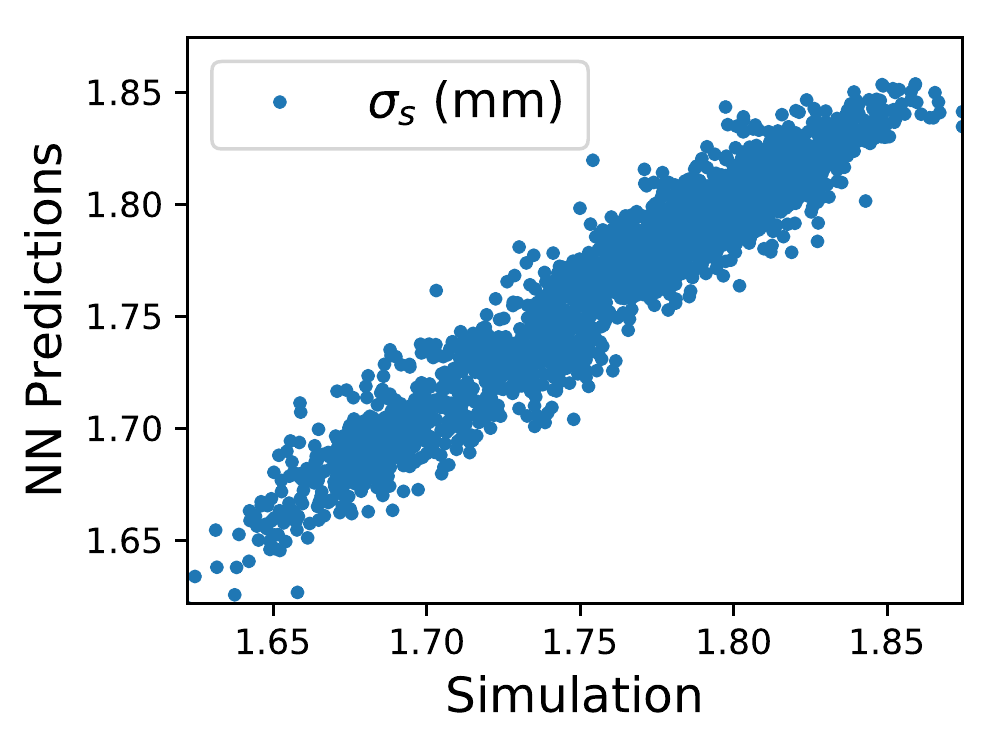}
	\includegraphics[width=0.3\textwidth]{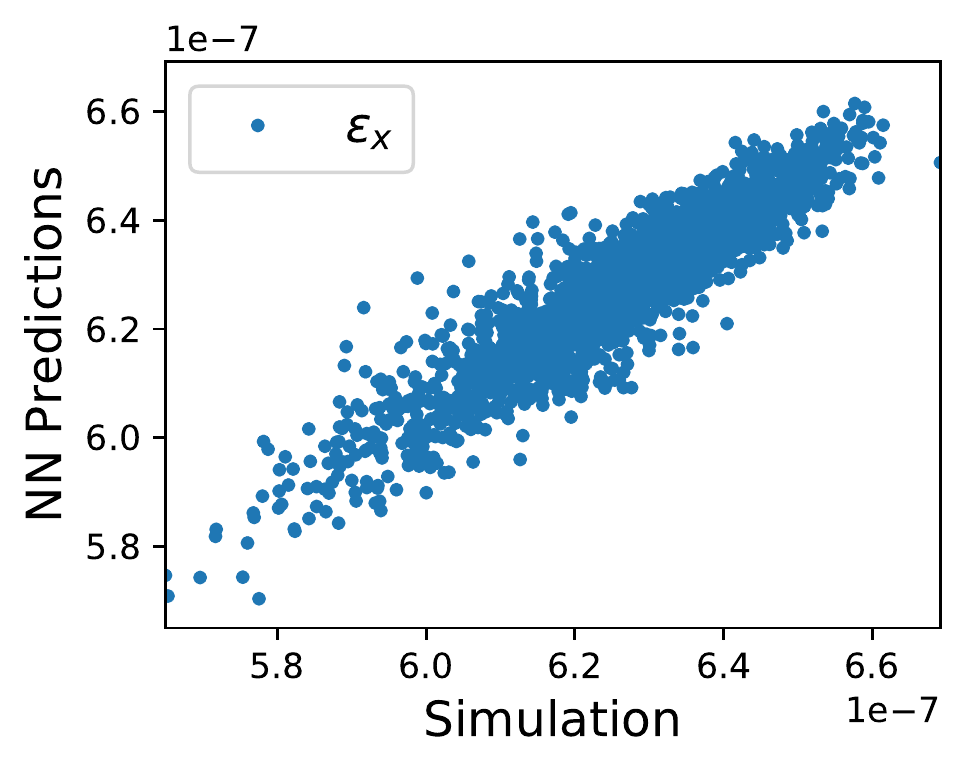}
	\includegraphics[width=0.3\textwidth]{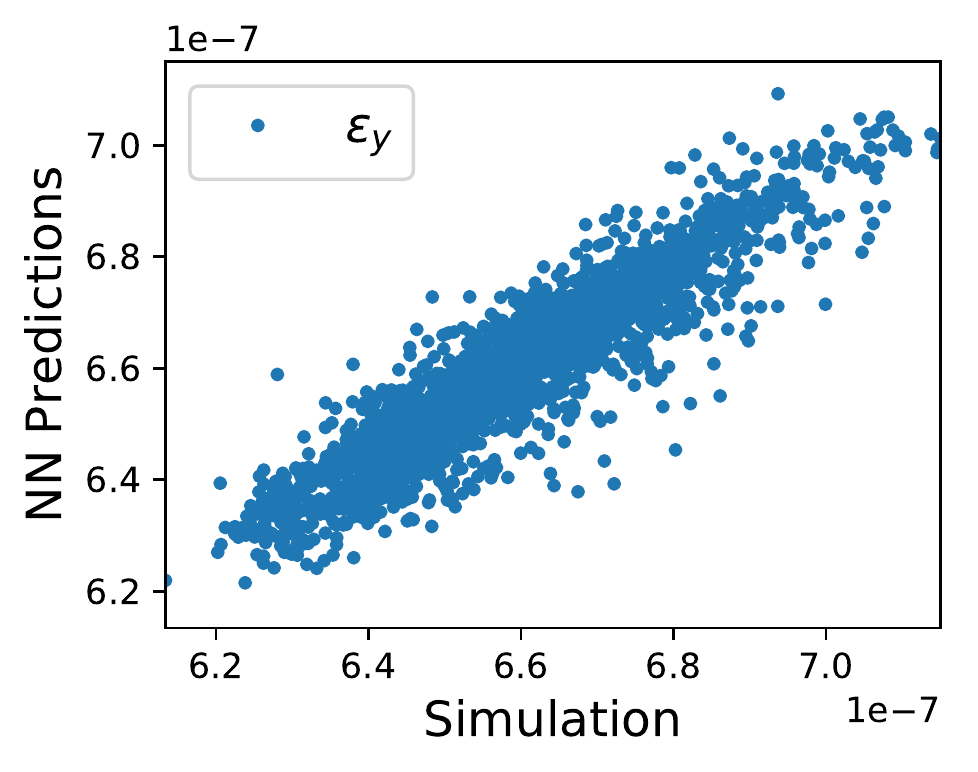}
	\includegraphics[width=0.3\textwidth]{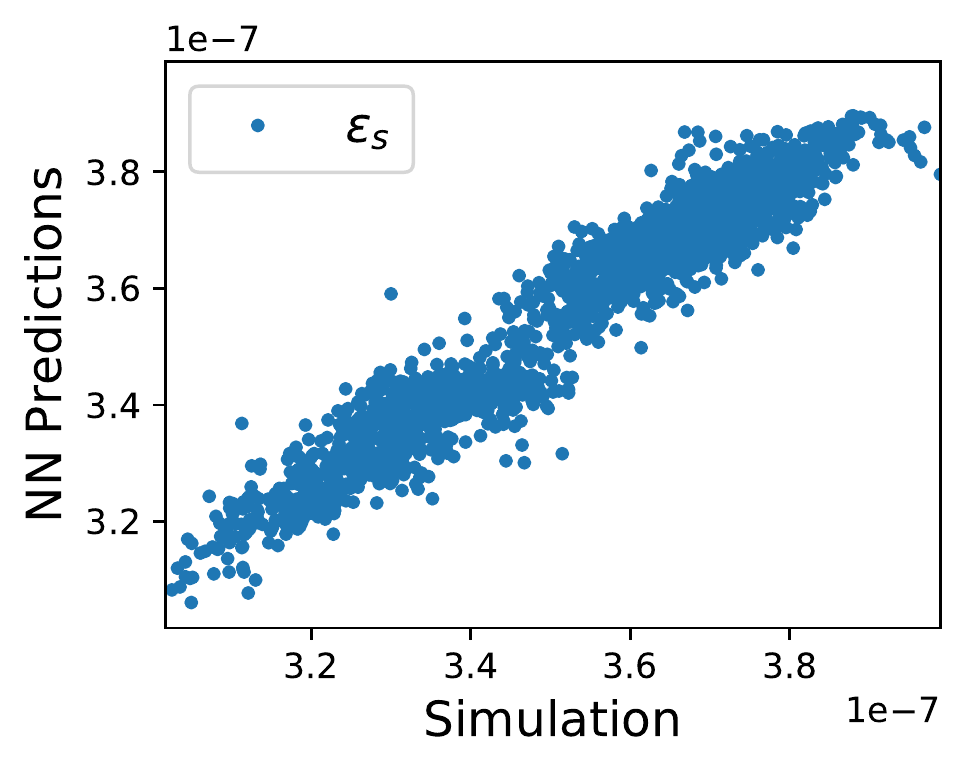}
	\includegraphics[width=0.3\textwidth]{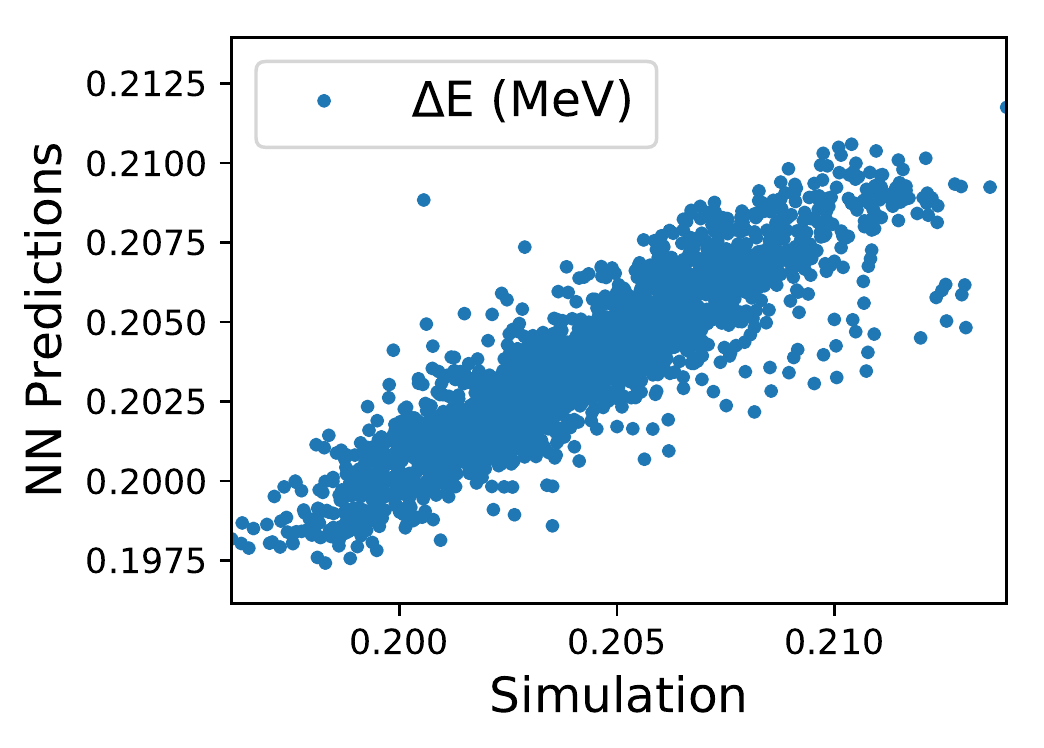}
	\includegraphics[width=0.3\textwidth]{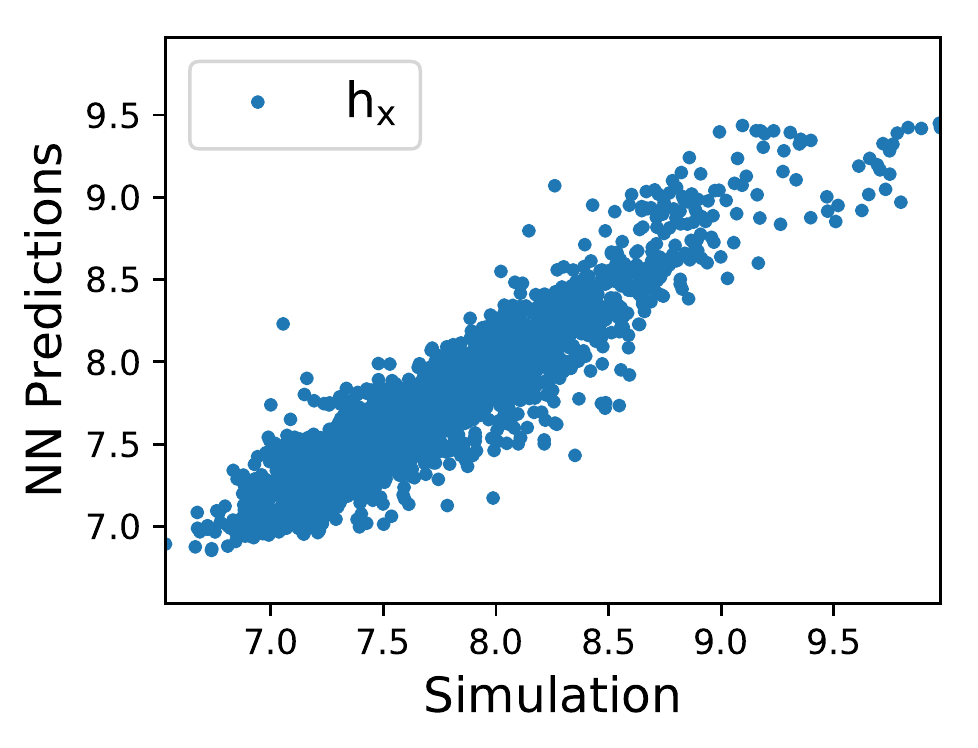}
	\includegraphics[width=0.3\textwidth]{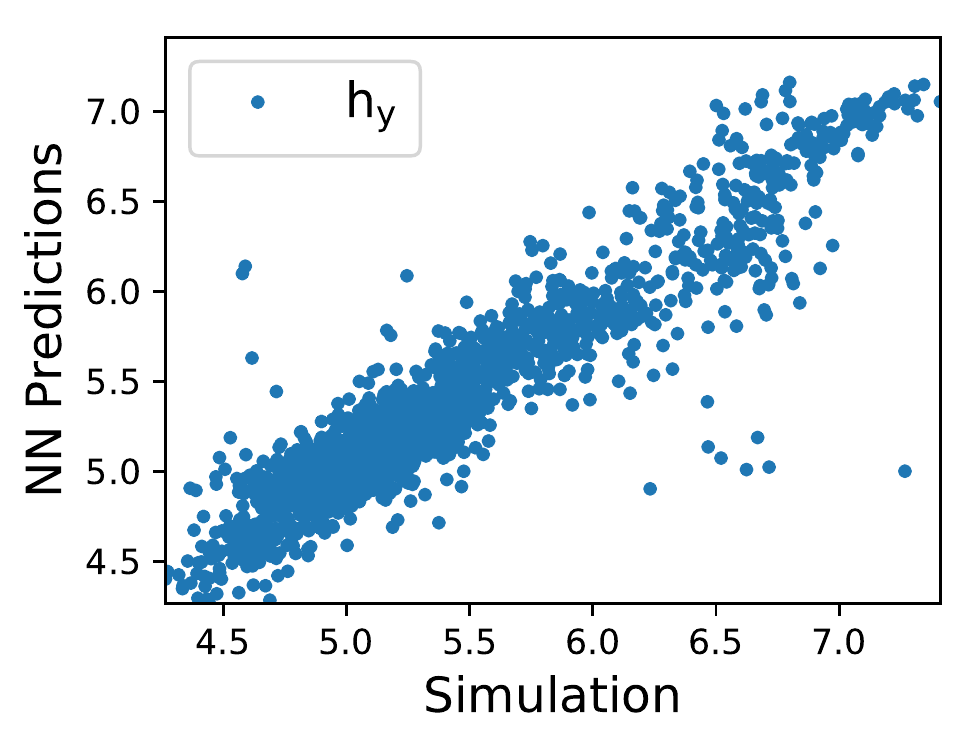}
	\includegraphics[width=0.3\textwidth]{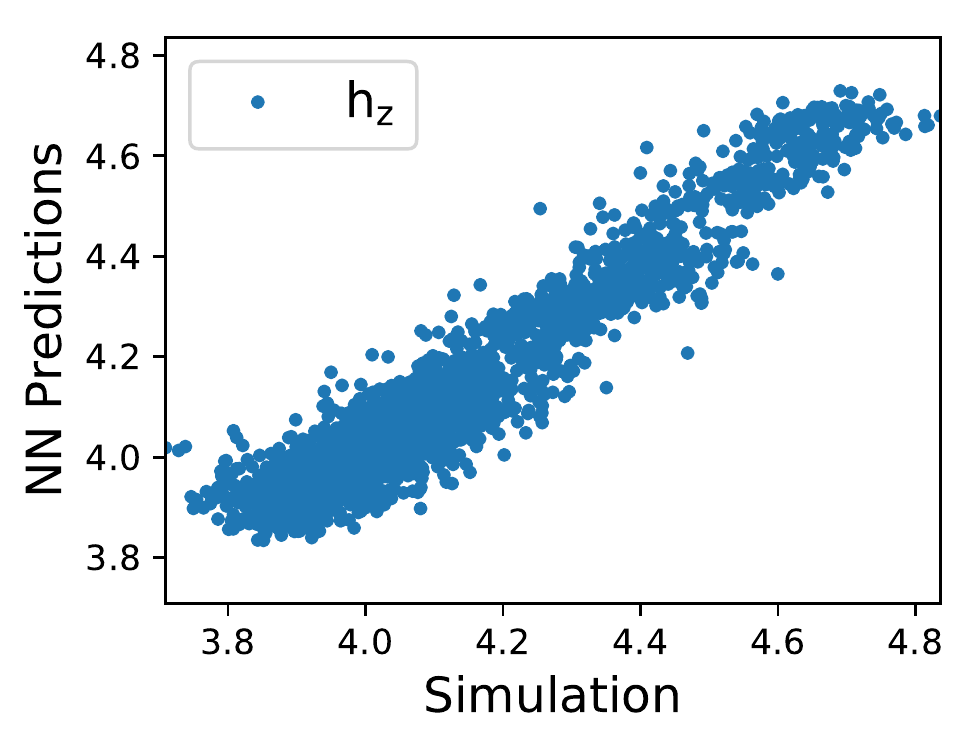}}

	\caption{Example predictions from IsoDAR model trained on 12 outputs vs. true values. Perfect prediction would correspond to a straight diagonal line. \label{fig: pareto_isodar1}}
\end{figure}

\begin{figure}[ht!]
	\centering
	{\includegraphics[width=0.3\textwidth]{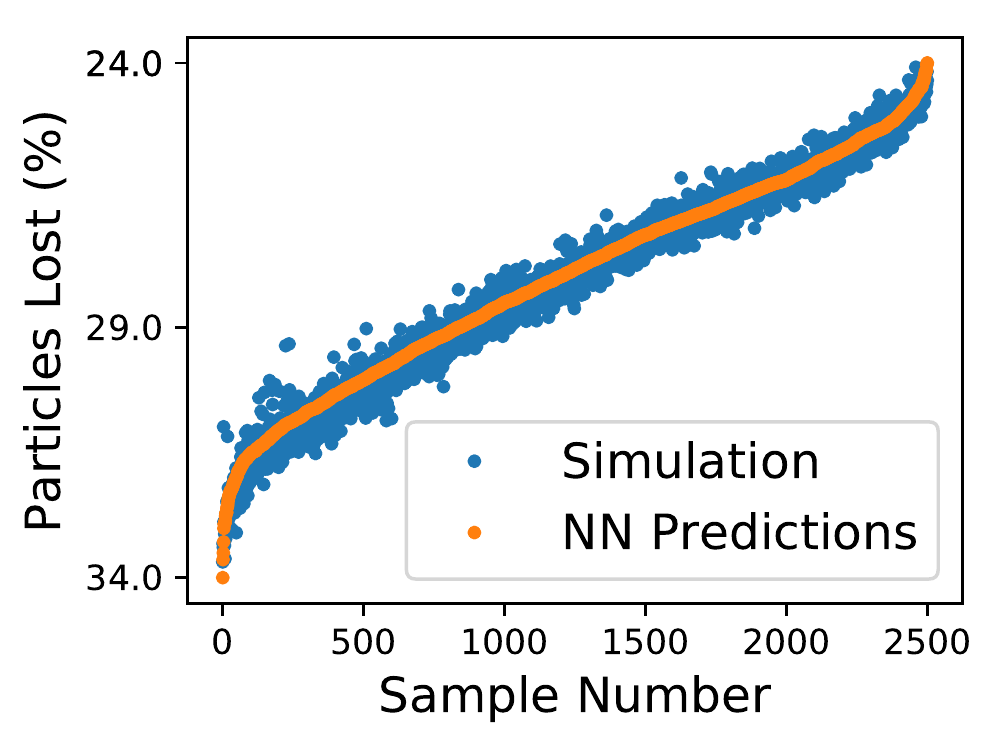}
	\includegraphics[width=0.3\textwidth]{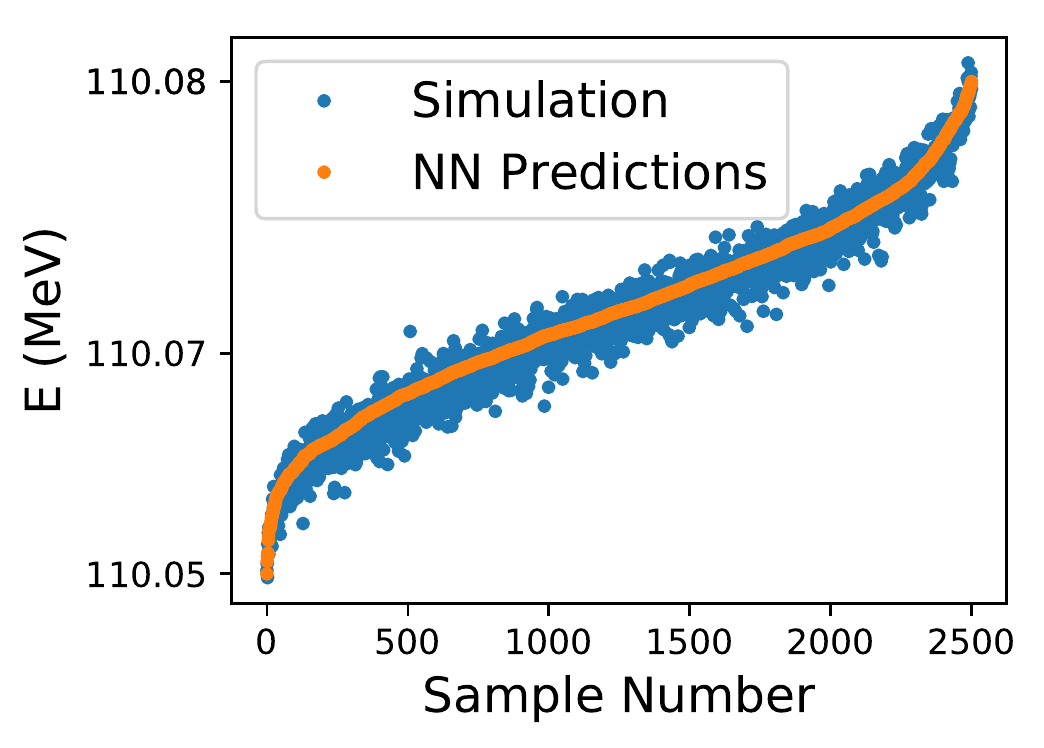}
	\includegraphics[width=0.3\textwidth]{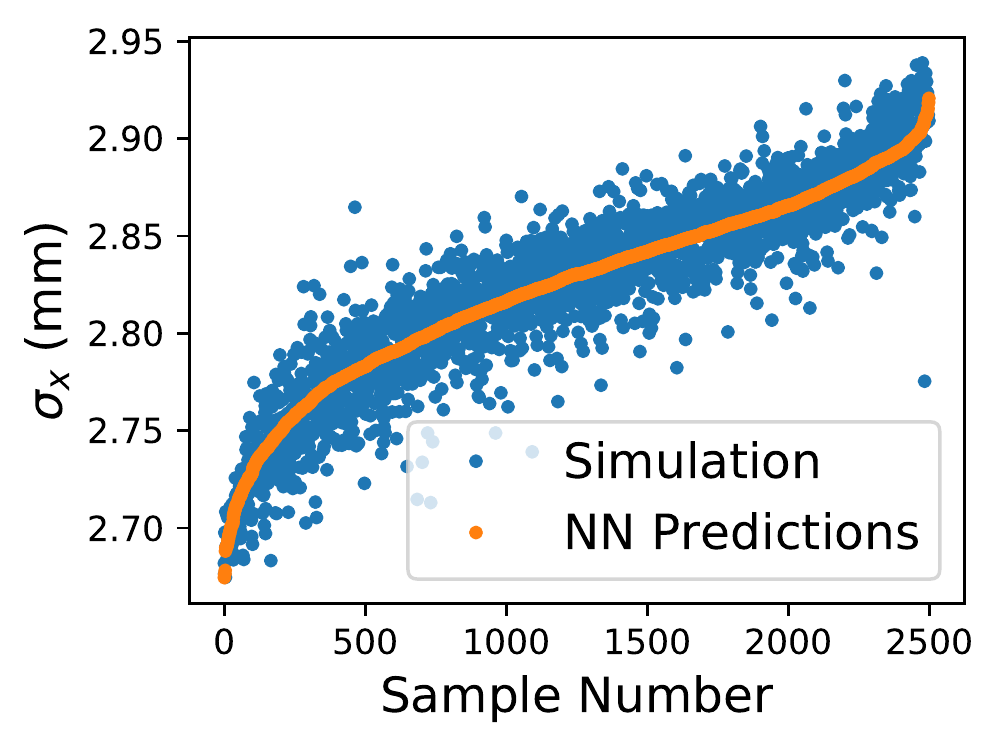}
	\includegraphics[width=0.3\textwidth]{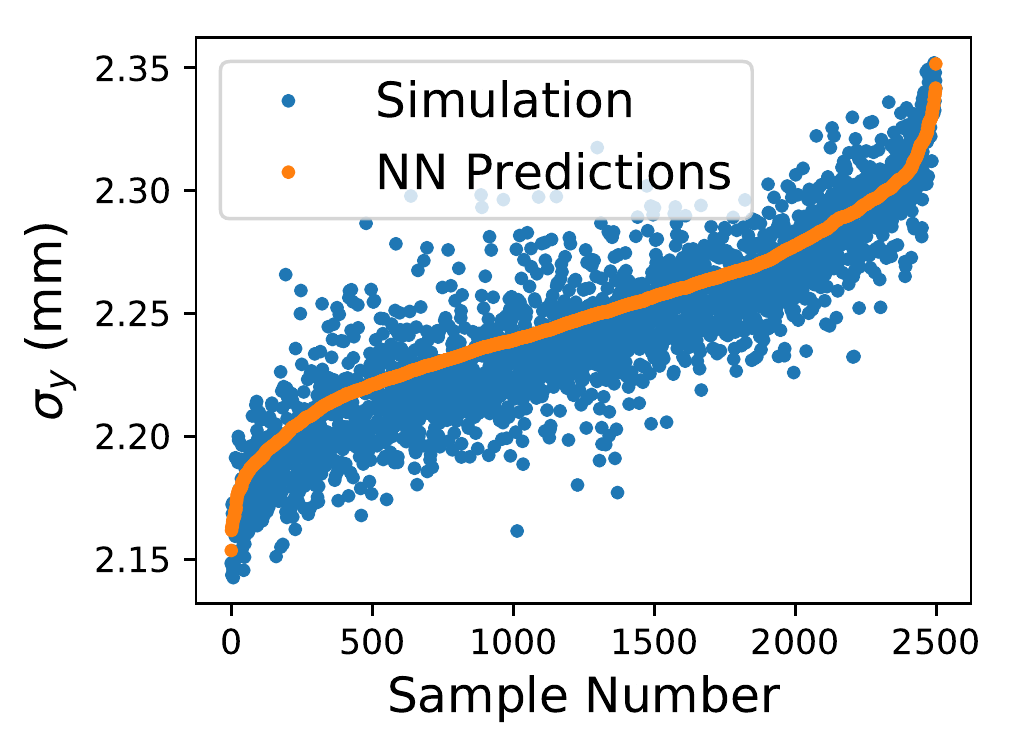}
	\includegraphics[width=0.3\textwidth]{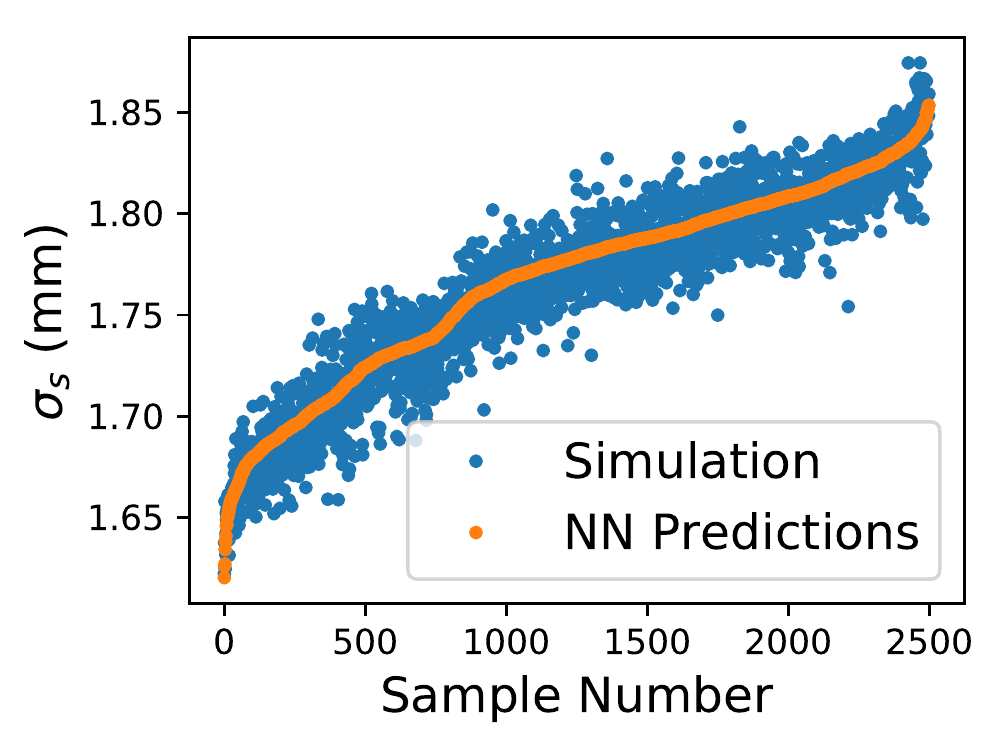}
	\includegraphics[width=0.3\textwidth]{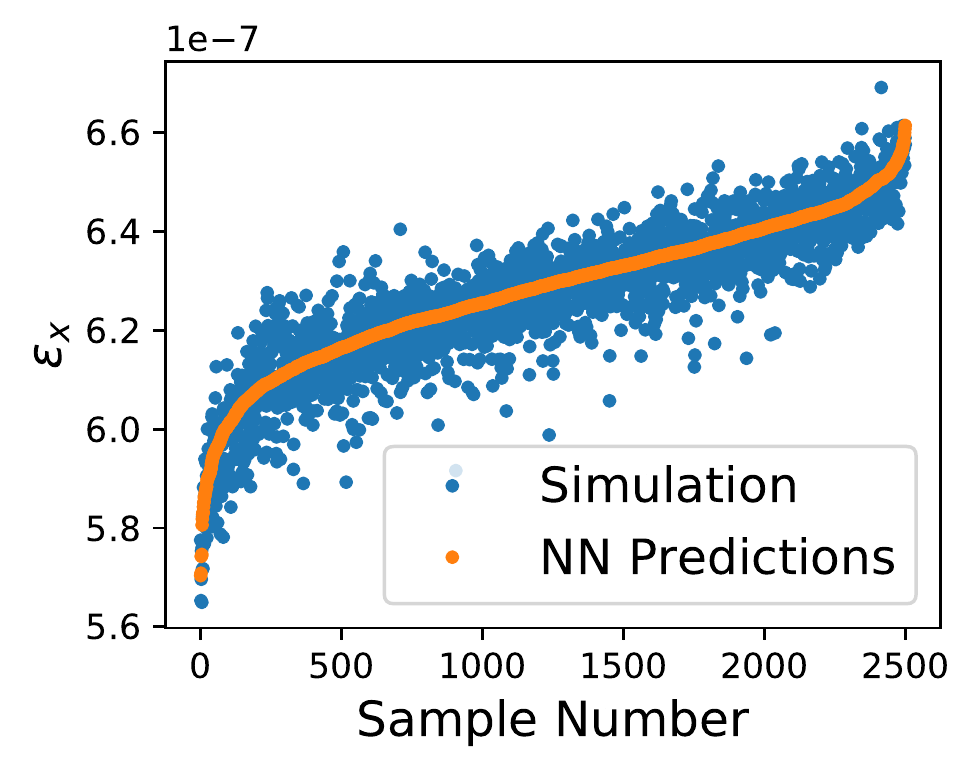}
	\includegraphics[width=0.3\textwidth]{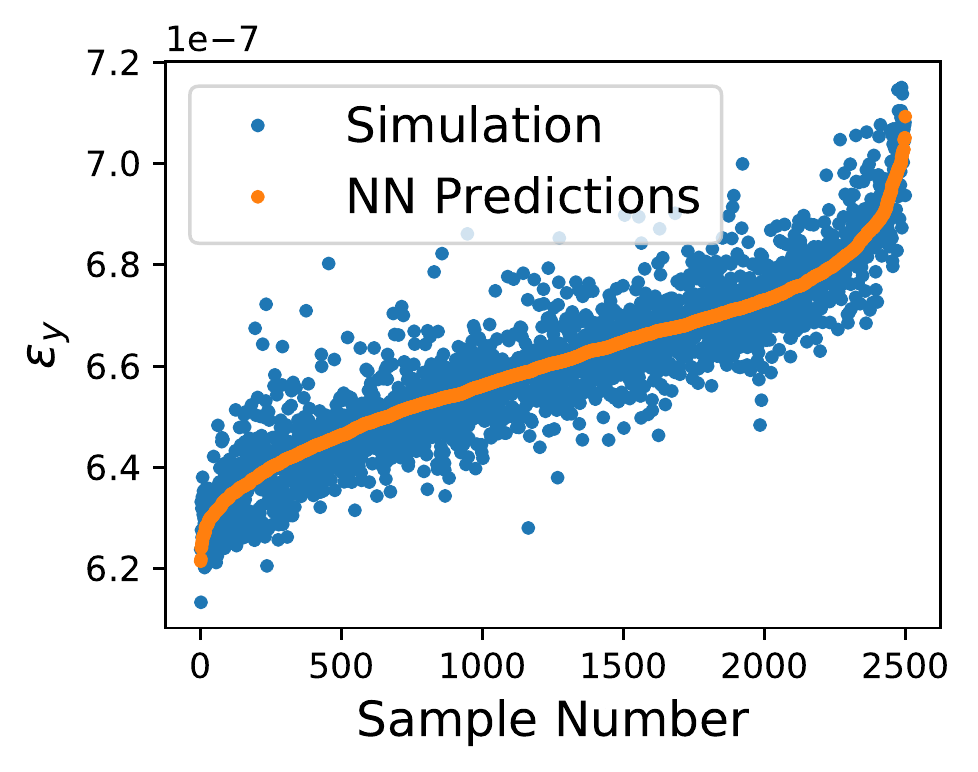}
	\includegraphics[width=0.3\textwidth]{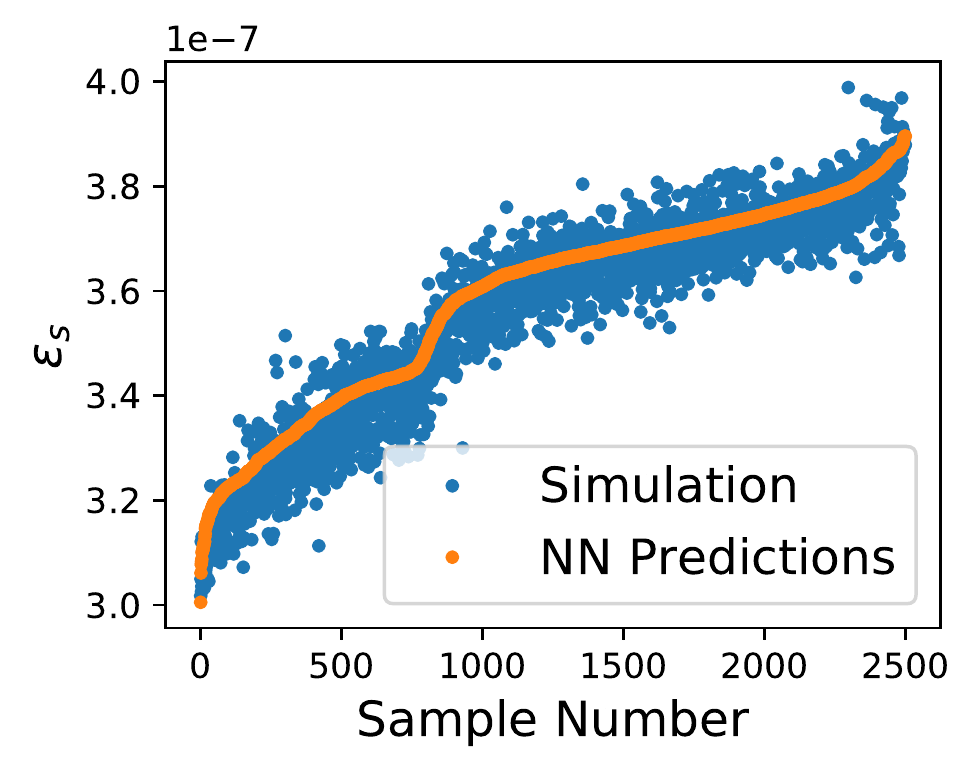}
	\includegraphics[width=0.3\textwidth]{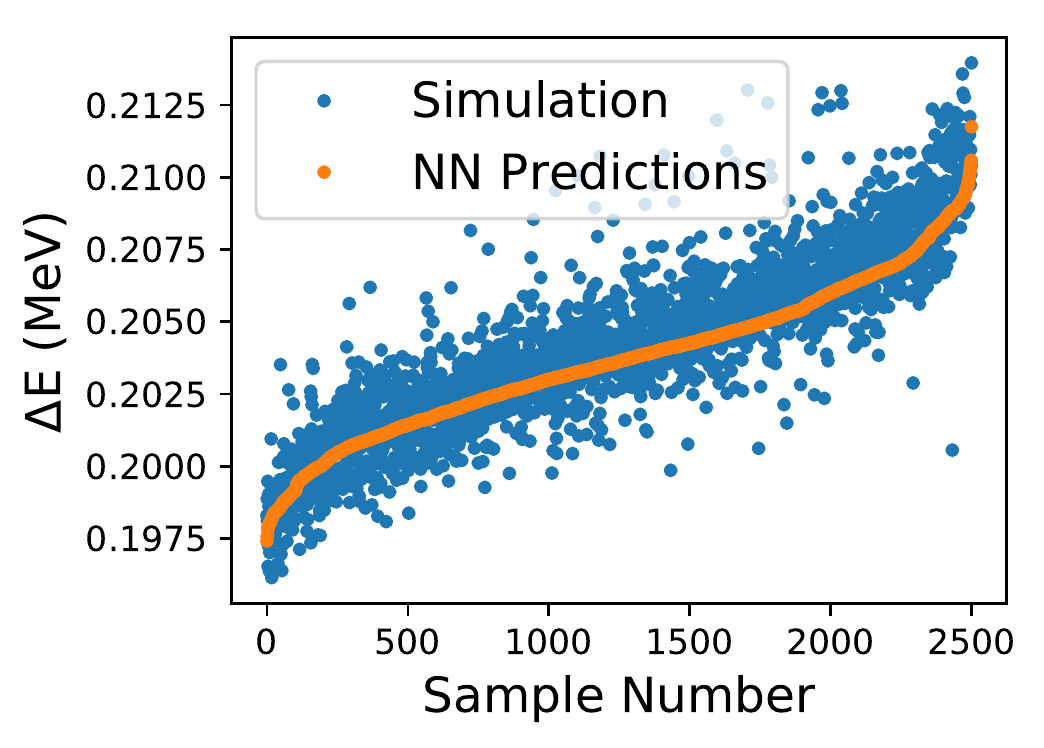}
	\includegraphics[width=0.3\textwidth]{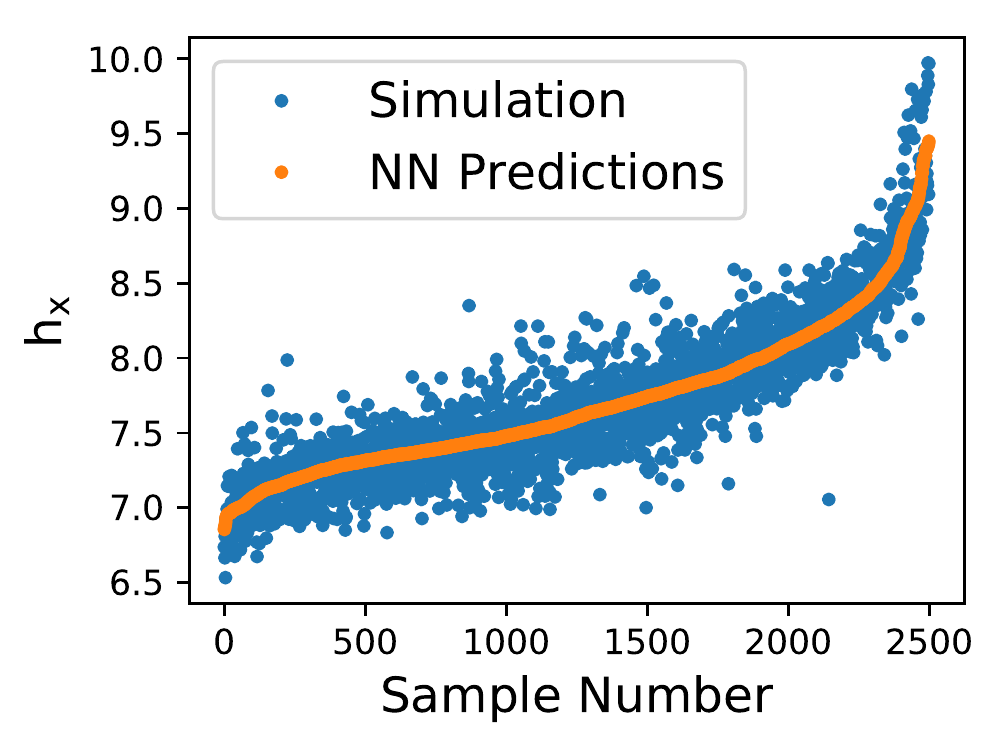}
	\includegraphics[width=0.3\textwidth]{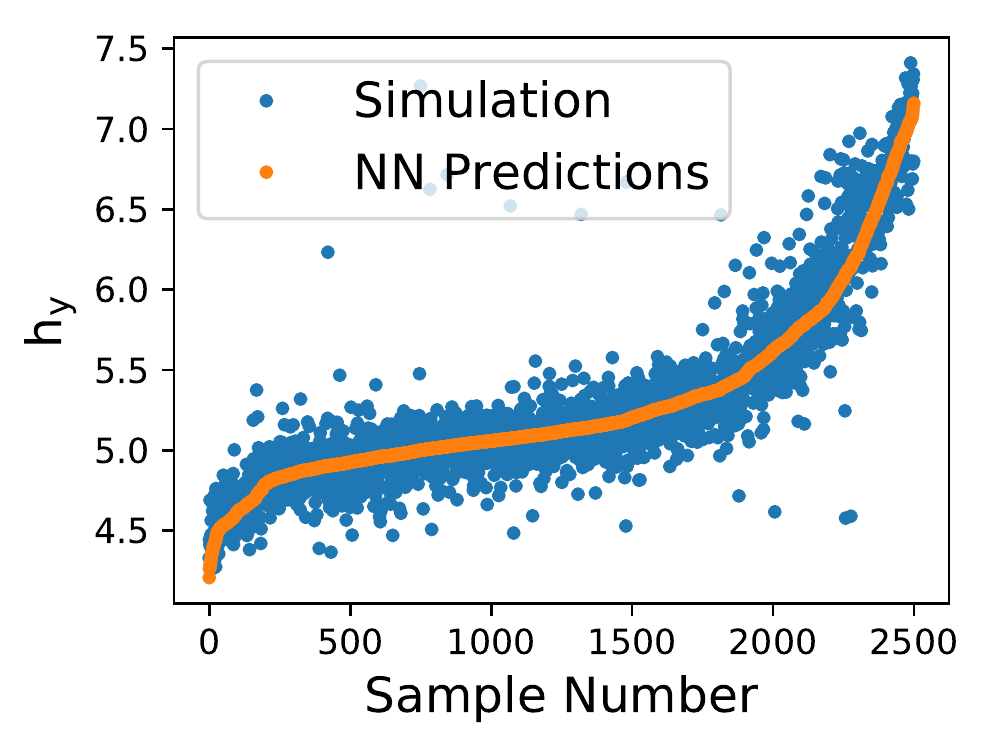}
	\includegraphics[width=0.3\textwidth]{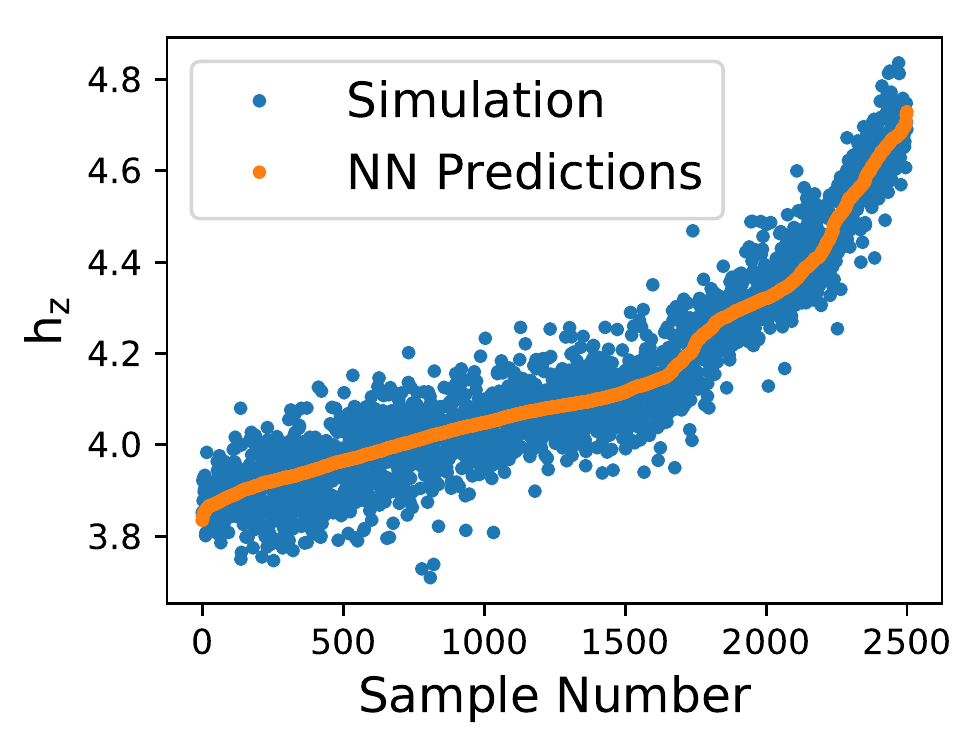}}

	\caption{Example predictions from IsoDAR model trained on 12 outputs and corresponding true values. \label{fig: pareto_isodar2}}
\end{figure}

In addition, we obtain a surrogate model that is $\mathcal{O}(10^{7})$ times faster to execute than the original physics simulation and can predict the 12 beam parameter outputs with  reasonable accuracy.  To illustrate this, in \figref{fig: pareto_isodar1} and \figref{fig: pareto_isodar2}, we show some examples of the prediction performance of the IsoDAR model on the random sample dataset. We show the prediction on the raw sorted values and plot the expected vs. predicted values against one another. The outliers in the dataset also illustrate the need for uncertainty estimates along with the prediction.

\clearpage
\section* {Conclusions and Discussion}

In practice, \psims\ of particle accelerators are often too computationally expensive for full exploration of the parameter space during optimization. The computational expense also limits (and in many cases prohibits) their use during machine operation to aid in prediction and control. In this work, we have shown that machine learning (and NNs in particular) can be used to obtain a fast-executing representation of computationally intensive accelerator physics simulations, and these models can be reliably used in multi-objective optimization. We have also shown that in some cases relatively little data is needed to achieve a high degree of fidelity relative to the original \psim.


The NN surrogate models are more accurate than the simplified physics models that are presently used when fast execution is needed (e.g., fast envelope codes, simple tracking codes used with a small number of particles, etc). This approach thus provides one avenue toward creating fast-executing representations of high-fidelity \psims\ for use in machine operation, which would not otherwise be possible. It also enables faster start-to-end optimization, which could in turn help facilitate more extensive offline design optimization and experiment planning, as well as aid the prototyping of novel accelerator operating modes. Finally, these models can be used to quickly prototype new optimization algorithms. 

Our results also suggest a new procedure for doing \moo\ optimization of particle accelerators. Instead of running a \moo\ with a  \psim, one can run a small random sample spanning the parameter space, train a surrogate model on this sample, and run a full \moo\ using the surrogate model very quickly. The estimated Pareto points obtained from the surrogate model could then be used as an initial population in a subsequent short \moo\ with the \psim\ to verify and fine-tune the result (see \figref{fig: new_meth}). The process can also be repeated until convergence to a good solution is reached (i.e. using iterative retraining and re-optimizing). We found that optimization with the NN model requires substantially fewer simulation evaluations than a purely simulation-based optimization (e.g., 132 times fewer simulations with one random sample in the case of the \SI{40}{nC} setup for the AWA, with indications that this could be further reduced to 330--550 times fewer simulations with iterative retraining). The computational cost of the parameter space exploration with the GA  before it gets close to convergence is high. Using the NN model enables one to skip these early stages of convergence and form a model that can be used to interpolate across the parameter space. Assuming the physical function to be modeled is smooth and that the NN model is learning a good representation of the underlying physical system, running the \moo\ with the NN model can thus provide an estimate of the Pareto front in a fraction of the time needed to run the \moo\ on the physics simulation. This opens up the possibility to do more extensive optimization than might otherwise have been feasible using computationally intensive \psims\ alone. Using NN models to aid the optimization process could be useful in cases where one cannot in practice run a \moo\ for a sufficient number of generations or with a large enough population to converge. This can be the case when HPC resources are limited and/or the optimization problem is too high-dimensional or computationally intensive. Related to this, a larger generation and population size can be used with the NN than would normally be feasible with the physics simulation, thus potentially enabling better solutions to be found during the optimization process (e.g. in contrast to doing only a short GA over a limited range with the physics simulation). Overall, while training a NN as an intermediate step in the optimization process may seem cumbersome, GAs do have their own hyperparameters that need to be tuned (e.g., population size, number of generations, crossover and mutation probabilities). The risk of wasting computational resources while tuning these parameters is high, and in practice accelerator physicists often just pick a number of generations to run based on experience.

To move from proof-of-concept demonstrations to regular deployment in accelerator applications, several areas of future work are apparent. These include addressing how best to scale these methods up to larger or more computationally intensive systems, how to obtain accurate estimates of prediction uncertainty, how best to combine simulation and measured data (i.e. to make efficient use of both, and to improve the model accuracy with respect to the real machine), how best to incorporate prior physics knowledge into these models, and how best to account for machine drift and keep the model updated over time. We discuss these in more detail below.

\textbf{Incorporation into online modeling and model-based control:} First, one can begin to incorporate these surrogate models directly in machine operation. In some cases, high-fidelity simulations have been made to match the machine very closely and can be used to provide suggested machine settings (for a comprehensive example, see~\cite{cornell_injector}). If the simulation matches the machine closely enough, the ML model trained on the simulations can immediately be used to aid operation.  Machine operators could use these models to check the potential impact of setting changes before trying them out on the actual machine, or to assess a new course of action as goals change during an operating shift. These models could also be used as a diagnostic tool to provide predictions about un-measured beam parameters (i.e., as a virtual diagnostic~\cite{mlwhitepaper,aedsurro3, lps_vd}) or to flag when the system has changed substantially (i.e., model-based anomaly detection). They could also be exploited in model-based control and model-guided optimization routines (i.e., using the model to help guide the search for optimal settings, as is done in model predictive control and Bayesian optimization ~\cite{Rasmussen2006}). These models can also be used to provide a warm start to a local, feedback-based optimization algorithm (i.e. by reading unvaried inputs from the machine to get an estimate of the present system state, running an optimization algorithm on the model for the variable settings to get an initial solution, and then refining this with online feedback with the machine directly). This is similar to the warm start approach described and tested in~\cite{lps_inv_es,aedsurro1, aedsurro2,aedsurro4,mlwhitepaper}, but it uses optimization around the forward model to obtain the initial suggested settings rather than using an inverse model. The advantage of the combined approach is that it can help compensate for inevitable discrepancies between the model and the machine (e.g., due to drift, hidden variables, etc.) without necessitating retraining, and the suggested settings from the model only need to be close enough to the basin of a good minimum to allow a local optimizer to converge. Forward models can also be used in the development of inverse models, which provide a direct map from desired beam outputs to suggested settings. In contrast to relying on local optimization methods or hand tuning only, these methods could be exploited to reduce the time spent switching between user requests. However, all of these possible applications for deployment in online modeling need to be explored in practice.
  
\textbf{Updating models with measured data:} Second, getting the physics simulation to match the measured machine behavior can be very difficult and usually requires substantial effort. As a result, many accelerator facilities do not prioritize creating accurate physics simulations (particularly since high-fidelity physics simulations could not be used directly in operation anyway, at least not prior to the introduction of the approach discussed in this paper). With the NN surrogate model, one can instead update the model learned in simulation with measurements from the machine to account for deviations between the simulation and the real machine behavior. This was shown to be viable in~\cite{aedsurro3}, but more rigorous study is needed to address how best to preserve information gained from the simulation while also updating the model with respect to measured data. By creating ML models that are trained at least in part on measured data, subtle statistical correlations across the machine that may otherwise go unnoticed and unutilized by human operators can be exploited. Training on simulations prior to this reduces the need to rely only on machine time and data available in the archive, thus potentially enabling regions of parameter space that otherwise would go unseen to be included. For some use cases, such as identifying the source of discrepancies between the machine and the simulation, it may also be more useful to learn to predict the error between the measured data and the model that was trained in simulation rather than updating the model directly. Finding ways to effectively combine measured and simulated data is an important direction of future work.

\textbf{Accounting for drift and unseen operating conditions:} Third, another challenge concerns how best to update the models with measured data in a reliable fashion to maintain prediction accuracy in the face of long-term drift in the machine response over time and exposure to new regions of parameter space. The first step of this is determining when the model has moved outside its range of validity relative to the present machine operating condition. Raw prediction errors for available signals or uncertainty predictions could be used to help decide when the model needs to be retrained and how much to trust a given prediction (e.g. to help decide whether to use that prediction in control or analysis, or rely on alternative methods instead). Then the model must be updated in a reliable fashion over time without manual intervention. For many applications, one cannot simply update the model automatically when a high error is observed, because this could be due to anomalous conditions (which may or may not already be automatically flagged as such). In addition, for accelerators that frequently switch operating conditions (e.g. beam energy), constant retraining when a high prediction error is observed could result in a loss of valuable information about previously-visited machine states. This is due to the well-known problem of ``catastrophic forgetting'' in NNs ~\cite{cf1,cf2,cf3}, and it was also observed in practice in early tests at the FAST injector ~\cite{privcom2} when attempting to do online retraining of the NN model described in ~\cite{aedmpc, nnsurvey} for different RF power levels. If one switches between configurations frequently, balancing the amount of information preserved from previously-visited states with the amount of information from new states will be critically important. Rather than retraining the entire NN over time, another avenue would be to take the approach used in ~\cite{virtual_accel}, where a local optimization algorithm is used to continuously update free parameters or unknowns in a simplified physics model so that it closely matches the observed beam. A similar approach could be used with an ML model instead of a simplified physics simulation. All in all, finding good strategies for updating the model or otherwise accounting for drift while preserving previously-learned behavior from different operating configurations will be critical. Depending on the particular details of the accelerator (e.g. how much drift there is from day-to-day, how much noise there is (e.g. jitter of the beam from pulse-to-pulse), to what extent does the diagnostic information capture the variables that affect the machine behavior), the strategies to do this most effectively will vary.

\textbf{Inclusion of model uncertainty:} Fourth, another challenge concerns how to obtain an accurate measure of the prediction uncertainty, particularly for NN-based models. Uncertainty predictions can be used to decide when the model needs to be re-trained, or when the model should not be relied upon. Although this is relevant to both online and offline applications, it is particularly important for online deployment or cases where measured data is used in training. This is because the input-output relationships of operational accelerators are subject to a variety of sources of uncertainty, including noise, intermittent anomalous conditions (e.g. due to equipment failures), drift over time, and the influence of unobserved variables. In the context of optimization, uncertainty predictions can also be used to help choose subsequent points to evaluate. For example, one may want to assess the expected benefit of running a computationally expensive simulation at a given operating point to fill out the parameter space of a model, or one may want to weigh the risk of moving to a new operating point on the live machine against the likely performance improvement from that new operating point. While this approach is well-established for Bayesian optimization techniques (especially using Gaussian Process models), obtaining reliable uncertainty estimates from NNs is an open area of research ~\cite{yar,ICML,bnn1}. Developing expressive models that also include uncertainty estimates (e.g., model ensembles, Bayesian NNs, Gaussian Process models with NN-based kernels \cite{wilson:dkl}, deep GPs \cite{Deep-GP}) is a reasonable next step. Critically, inclusion of reliable uncertainty estimates into NN models like the ones we show in this work would enable them to be used with Bayesian optimization. Bayesian optimization with Gaussian Process models has been successfully used in online optimization of operational accelerators ~\cite{bayesopt1, bayesopt2, bayesopt3}, and a pre-trained NN model covering a broad set of controllable variables could in principle be used in a similar fashion.

\textbf{Efficient sampling strategies:} Sampling of simulations to produce an initial model can itself be time-consuming in cases where the cost of obtaining a sample is extremely high or when many variables must be included. In some cases, the random sampling and iterative retraining strategies we used in this work may not be sufficient, and more intelligent sampling strategies will likely be needed in order to map the parameter space fully while minimizing the number of simulation evaluations. Even in the AWA case, we observed that for the random sample, some areas of the output space were  over-represented and others were under-represented. Bayesian optimization is one promising candidate that could be used to sample the parameter space more efficiently to produce a good model (i.e. by prioritizing sample sparsity). 

\textbf{Scaling to higher dimension and complexity:}  While NNs in principle can be used to model high-dimensional, complex systems, determining how best to scale this method to accelerator systems with a much greater number of input/output variables, wider ranges of variables, or more complex beam dynamics is an important question that will also need to be examined in future work. 

\textbf{Including prior physics information:} Present approaches discard our rich knowledge of accelerator physics. Methods should be developed which incorporate this physics knowledge into these models so that fewer training examples are required and improved performance in unseen regions of parameter space can be obtained. 

Especially as research in the above areas progresses, we anticipate that the approach presented in this work will be useful for a variety of applications, including design optimization, prototyping of optimization routines, offline experiment planning, and online optimization of machine settings. Overall, the approach is quite general, and as many beam dynamics problems in accelerators are similar, it is also reasonable to expect that these results could provide good starting points for applying this approach to other kinds of accelerator systems. For example, injector systems that are similar in scale and complexity to the AWA are extremely common, and the results we show in this work should provide good guidance for those wishing to use this method on similar components. The fact that we used a similar NN architecture to other injector modeling problems (e.g. ~\cite{FAST, aedsurro1, aedsurro2, aedsurro3, aedsurro4, privcom,ml4phys}) also hints at the possibility of doing transfer learning between models (e.g., training on one injector system and then re-using the model with small updates for a similar injector system). While the IsoDAR cyclotron is a more unique design, the performance of the ML approach on that case shows it can also be used in cases with much more complicated beam dynamics.

\clearpage
\section*{Appendix} \label{sec: theory}

\subsection{Data Sets for the Surrogate Models}
We generated uniformly-distributed random samples from the \psim. For this we used the \opal-based interface for creating such data sets, which was developed in part to support this effort. This feature allows the submission of massively parallel jobs using an \opal\ input file. 

For the AWA, the randomly-varied inputs include the injector phase $\phi_1$ and gradient $G_1$, the linac cavity phase $\phi_2$ and gradient $G_2$, and two solenoid strengths $K_1$ and $K_2$. The output parameters are the transverse spot sizes $\sigma_x$ and $\sigma_y$, the longitudinal beam size $\sigma_s$, the transverse projected emittances values $\varepsilon_x$ and $\varepsilon_y$, the longitudinal projected emittance $\varepsilon_s$, and the energy spread $\Delta$$E$. The input variable ranges are informed by the operating ranges at the AWA (see \tabref{tab: desparam}). Random samples for two bunch charges were generated (\SI{1}{nC} and \SI{40}{nC}, with the corresponding laser radius being \SI{2}{mm} and \SI{9}{mm}). In the \SI{1}{nC} case, we generated 70k samples, and in the \SI{40}{nC} case, we generated 80k samples. However, we only use a small subset of these during training of the models.

\begin{table}[hb!]
\centering
\caption{Range of the AWA Input Variables. \label{tab: desparam}}
\medskip
\begin{tabular}{lcccl}
\hline
Name & Abbreviation & Min Value & Max Value  & Unit\\
\hline
Solenoid 1 Strength           	  	 & $K_1$ 	& 400 & 550 & m$^{-1}$ \\
Solenoid 2 Strength           	  	 & $K_2$  	& 180 & 280 & m$^{-1}$ \\
Injector Phase              			 & $\phi_1$ 	& -10 & 0   & deg\\
Cavity Phase                			 & $\phi_2$ 	& -10 & 0   & deg \\
Injector Accelerating Gradient  & $G_1$ 	& 60  & 75  & MVm$^{-1}$ \\
Cavity Accelerating Gradient    & $G_2$ 	& 15  & 25  & MVm$^{-1}$\\
\hline
\end{tabular}
\end{table}

For the IsoDAR cyclotron, the input parameters varied are the initial beam current $I_{inj}$ and four collimator settings 
$C_1$--$C_4$. The ranges of the collimators and the beam current are determined by technical design considerations. 
A random sample of 2500 points was generated for the initial training data set. For retraining, 100 points from the previous estimated  Pareto front were selected randomly to add to the training set.

\begin{table}[hb!]
\centering
\caption{Range of the IsoDAR Input Variables. \label{tab: desparam}}
\medskip
\begin{tabular}{lcccl}
\hline
Name & Abbreviation & Min Value & Max Value  & Unit\\
\hline
Initial Beam Current           	  	 & $I_{inj}$ 	& 5.5 & 7.5  & mA \\
Collimator 1          	  	 & $C_1$  	& 2.37 & 2.63 & unitless \\
Collimator 2         			 & $C_2$ 	& 2.37 & 2.63 & unitless\\
Collimator 3           			 & $C_3$ 	& 7.60 & 8.40   & unitless \\
Collimator 4  & $C_4$ 	& 7.60 & 8.40  & unitless \\

\hline
\end{tabular}
\end{table}

\begin{table}[h!]
\setlength{\tabcolsep}{10pt} 
\renewcommand{\arraystretch}{1.25} 
\centering
\caption{Definitions of beam parameters referred to in the text.  \label{tab: beamparam}}
\medskip
\begin{tabular}{lccp{58mm}}
\hline
Name & Symbol & Unit & Definition  \\
\hline
Position                  & $x, y, z$                    &      m     & Position of each particle \\
Momentum                 & $p_x, p_y, p_z$                  &     1      &  $\gamma \beta_x, \gamma \beta_y, \gamma \beta_z$ \\
Normalized emittance       & $\epsilon_x, \epsilon_y$ 	        & m-rad   & $\sqrt{\left<x^2\right> \cdot \left<p_x^2\right> - \left<x\cdot p_x\right>^2}$, similar for $y$.\\
Bunch length              & $\sigma_z$, $\sigma_s$          	& m        & rms bunch size in $z$ or $s$ \\
Transverse beam size      & $\sigma_x$, $\sigma_y$ 	& m        & rms bunch size in $x$ or $y$\\
Energy                    & $E$ 	                & MeV       & Mean bunch energy \\
Energy Spread             & $\Delta E$ 	                & MeV       & Energy spread of the bunch \\
Halo                      & $h_x$ 	                & 1  & Particles outside the core, $\frac{\left<x^4\right>}{\left<x^2\right>^2}$ \\
Particles Lost            & $P_L$ & \% & {Percentage of initial particles that travel outside the accelerator aperture} \\
\hline
\end{tabular}
\end{table}

For running the GA on the AWA simulation, we choose an initial population of 656 individuals
and subsequently evolve the population over 200 generations, while retaining the same number of individuals in each generation. The following hyper-parameters were used: gene mutation probability $P_g=0.8$, mutation
probability $P_m = 0.8$, and recombination probability $P_r=0.2$. Specific descriptions of the hyperparameters can be found in Section~1.4.2 of the \opal\ manual~\cite{opal}. The specific values used in this case were chosen based on previous optimization work for the AWA that involved a hyperparameter scan~\cite{Neveu:2013ues}. The constraints of the MOO problem 
were set such that the variables stayed within the operating ranges of the AWA (see \tabref{tab: desparam}).
For each generation, the input and output parameters from the simulation are saved. In the \SI{1}{nC}  and \SI{40}{nC} cases respectively, 59,285 and 65,929 final samples were obtained. See ~\tabref{tab: computation_times} for an overview of the computational resources required to make the physics simulation data sets. 

Note that for the IsoDAR cyclotron, we do not run a GA on the physics simulation. This is because the IsoDAR simulation is so computationally expensive that we were unable to successfully run a multiobjective optimization on all parameters to convergence. 

The implementation of NSGA-II ~\cite{nsgaii} that is provided within \opal\ was used for the \moo\ with the physics simulation of the AWA. Details related to the algorithm can be found in~\cite{Yves1}.  The implementation of the GA used with \opal\ is slightly different than the implementation in DEAP (used with the ML models) because some of the adjustable hyperparameters are defined differently. Originally, we had intended to use DEAP for both the physics simulation and the ML model. However, we ran into practical limitations in being able to run the OPAL simulation in parallel using DEAP. OPAL had ostensibly the same algorithm programmed in a way that allowed efficient parallelism (making it feasible to run the GA on the physics simulation in a reasonable amount of time). We did not at the time have a way to run DEAP in parallel on the HPC systems that we had access to. Thus, we used the same nominal algorithm in each case (NSGA-II), but the implementation is slightly different in DEAP and OPAL. 

\begin{figure}[ht!]
	\centering
  	{\includegraphics[width=0.4\textwidth]{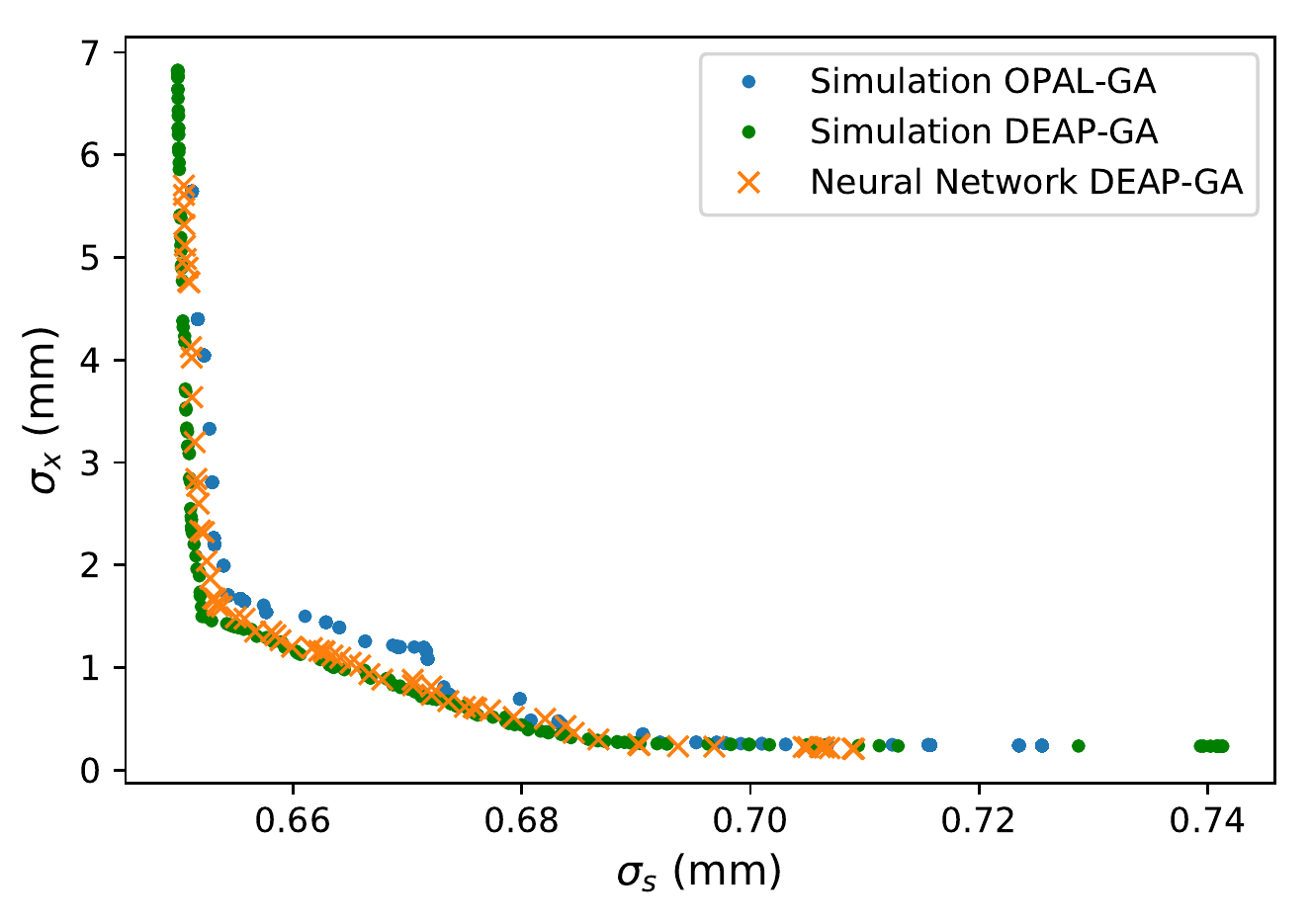}
	\includegraphics[width=0.4\textwidth]{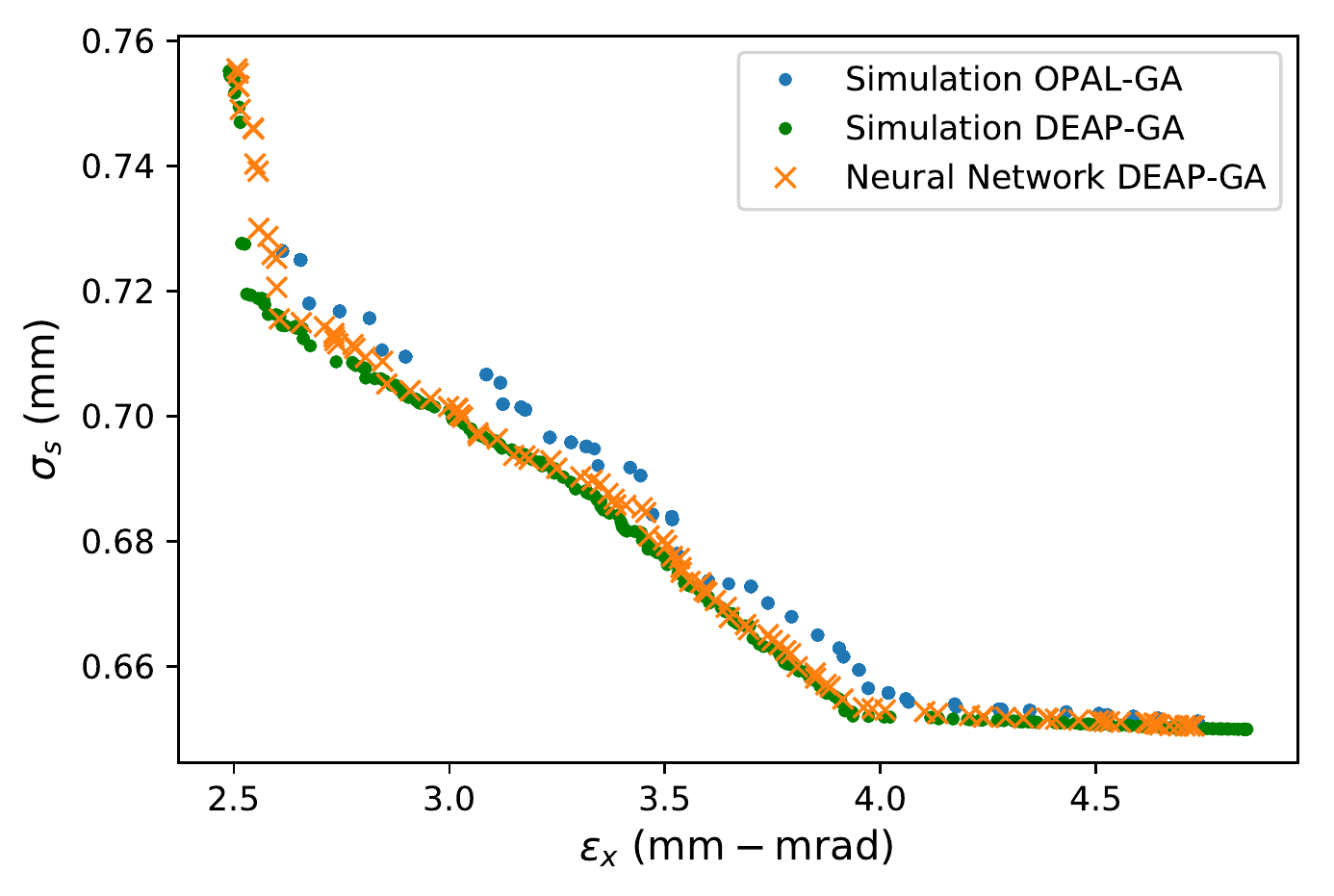}
	\includegraphics[width=0.4\textwidth]{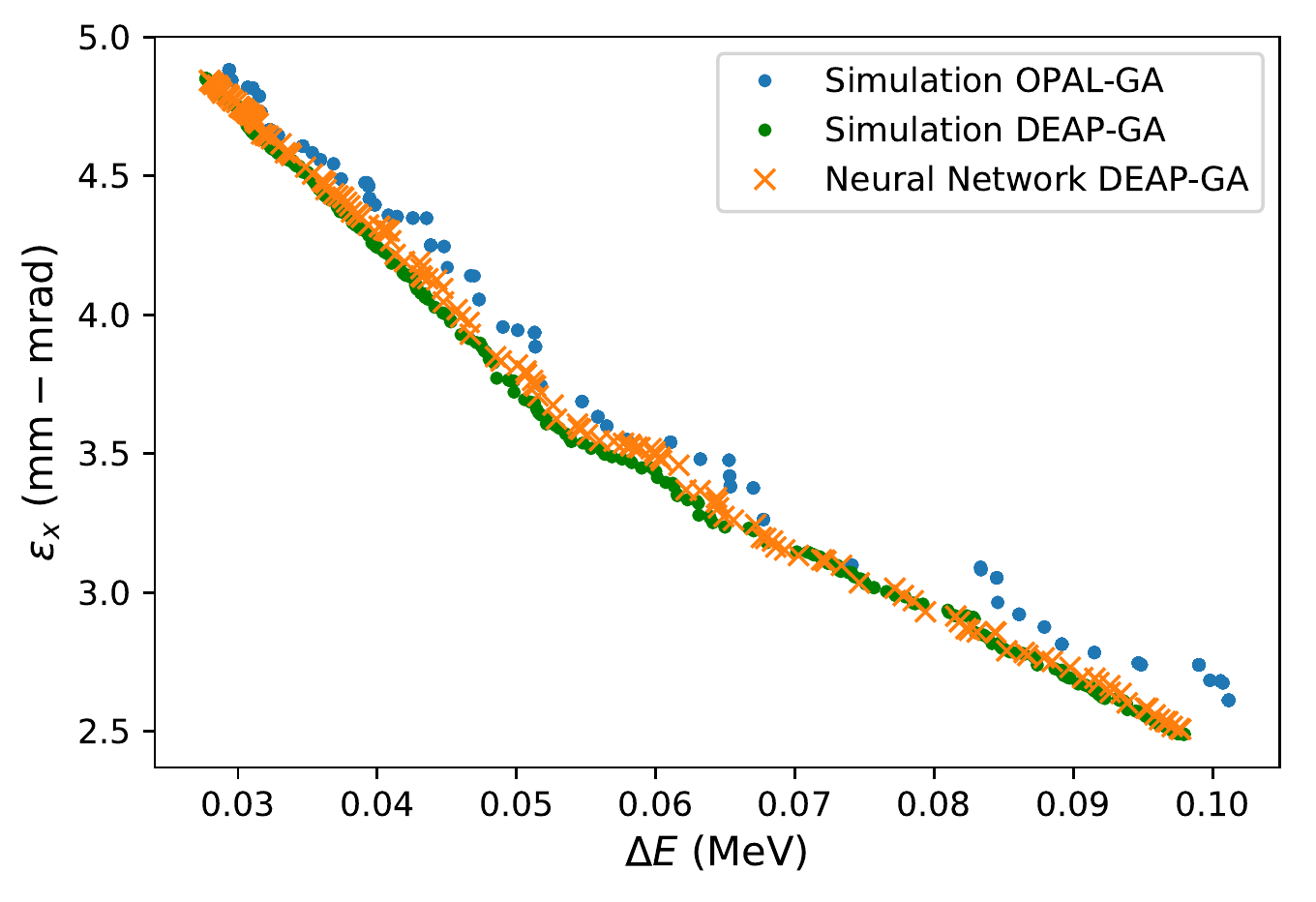}}
	
  	\caption{Comparison between estimated Pareto fronts obtained with the OPAL implementation of NSGA-II and the DEAP implementation of NSGA-II after 200 generations, with 656 individuals in each generation. We also show the result from running DEAP on the neural network surrogate model. We consider these to be in sufficiently close agreement; please refer to the text for more discussion. \label{fig: pareto_DEAPOPAL}} 
\end{figure}

After setting up libEnsemble ~\cite{libensemble} to do the parallel simulations in conjunction with DEAP, we were later able to do a direct comparison between DEAP's version of NSGA-II and OPAL's version by running both of these with the physics simulation. When comparing the two algorithms on \opal\ simulations of the AWA, the fronts produced by each method are in reasonable agreement, as shown in \figref{fig: pareto_DEAPOPAL}. The main source of the discrepancy between the OPAL-GA solution and the DEAP solution is the slight difference between the algorithm implementation. Note that this does not alter the conclusions of the paper. Specifically, two points need to be considered: (1) the solution from the NN model is in agreement with both the OPAL and DEAP solutions (which themselves have deviations between one another but are in general agreement) after running 200 generations with 656 individuals in each generation, and (2) the beam parameters found with the NN model are more optimal than are seen in the training set (which is a separate consideration from the DEAP or OPAL GA solutions).

\begin{table}[hb!]
\caption{Overview of the data sets from the physics simulations and the computational resources used for generating them. Note that we found that to make accurate surrogate models we needed many fewer randomly sampled points than were initially generated. For the GAs, 2624 cores were used. For the 1nC random sample, 16k cores were used, and for the 40nC random sample,  15k cores were used. \label{tab: computation_times}}
\begin{tabular*}{\columnwidth}{@{\extracolsep{\fill}}llcccc}
\toprule
Case		& Calculation 							& Core-hours		& Wall time (hours)			& Sim. evals.    \\
\midrule
AWA \SI{1}{nC}	 &Genetic Algorithm (200 gen)		& 43,500			& 16.56		& 65,929 \\
				&Random Sample						& 60,600				& 4.12		& 70,000 \\  
				&500 points 						& 283 				 	& 0.14	    & 500 \\ 
\midrule
AWA \SI{40}{nC} 	&Genetic Algorithm (200 gen)    &  95,000				& 36		& 65,928 \\
			&Random Sample					        & 115,700			& 7.23 		& 80,000 \\  
			&500 points 						    & 660				    & 0.33 		& 500 \\
\midrule
IsoDAR &Random Sample	                                           & 13,335       			 & 1.7       					 & 2,500\\
\bottomrule
\end{tabular*}
\end{table}


\subsection{OPAL Simulation}

\opal{} is an open-source, parallel library for electrostatic PIC simulations of charged particle accelerators. More details can be found in the OPAL manual (see~\cite{opal}). For simulations of the AWA, 3D space charge forces are calculated throughout the time-evolution of the beam, which is important for realistically capturing the nonlinear impact of the beam self-fields. The particle generation at the CsTe photo-cathode is modeled using an uniform emission model, 
assuming a planar ideal surface. The laser profile used for emission is uniform transversely and a flattop longitudinally, with Gaussian tails. Convergence studies were previous done to determine an appropriate time step,  mesh  size,  and  number  of  particles to use (here, $1\times10^{-11}$ seconds for the time step,  16$\times$16$\times$32 grid cells for the space charge mesh, and 10k macro-particles). 
The full-width-half-maximum of the laser in the longitudinal direction was \SI{6}{ps} for both cases.
The laser radius was set to \SI{2}{mm} for the \SI{1}{nC} simulations. 
Due to large nonlinear space charge forces at \SI{40}{nC}, the laser radius was increased to \SI{9}{mm}. 
These are typical operating conditions at the AWA. 

Field maps generated in POISSON~\cite{poisson} were used to model the solenoid magnets.
Two types of rf field maps were used to model the gun and accelerating cavities.
2D maps were generated in SUPERFISH~\cite{superfish} and used in the \SI{1}{nC} case.
3D maps were generated in ACE3P~\cite{ace3p} and used in the \SI{40}{nC} simulations.
While 3D field maps are computationally more expensive to evaluate, 
they are more accurate and capture asymmetries that are present in the AWA rf cavities.

The IsoDAR simulation is described in section II of \cite{isodarsim}. In addition, in this work we
added 4 collimators to clean up the beam (i.e.,\ reduce halo). The collimators are placed in the central region of 
the cyclotron where the energy is low and the activation is negligible.  

\subsection{Implementation of Machine Learning Based Surrogate Models}

The NNs were implemented in Keras~\cite{chollet2015keras}, with TensorFlow~\cite{tensorflow2015-whitepaper} as the backend. For general demonstration of the technique, we used a topology and set of hyperparameters that the authors had previously found to work well for similar problems in accelerators~\cite{FAST,aedsurro1,aedsurro2,aedsurro3,aedsurro4 }. This consisted of a fully-connected, feed-forward NN with four hidden layers, each with 20 nodes and hyperbolic tangent activation functions. No regularization penalties (e.g.,\ $L_{1}$ or $L_{2}$ norm) were used on the weights. The NNs were trained for 10k epochs with a batch size of 500 points. The Adam optimization algorithm~\cite{Kingma2014} was used for training, with an initial learning rate of 0.001 and hyperparameters $\beta_{1} = 0.9$, and $\beta_{2}=0.999$. For training, the random sample data was randomly split into training (60\%), validation (20\%), and testing (20\%) sets. All data sets were scaled to fit within an appropriate range. For example, in our case the data was scaled to be within the range of $[-1, 1]$. For the IsoDAR problem, the setup is the same, except we use a neural network with a slightly different number of nodes in each hidden layer: $10-20-20-15$ nodes in each layer respectively.

The surrogate model based on polynomial chaos expansion (PCE) is constructed
using the Uncertainty Quantification Toolkit (UQTk)~\cite{Debusschere2016}.\ This library provides
functionalities to perform an intrusive as well as a non-intrusive UQ in C++
and Python.\
In contrast to the projection method of~\cite{aa1}, we used the regression
method~\cite{doi:10.1137/S1064827503427741, SUDRET2008964} with Legendre polynomials, and we associate
a uniform distribution to all input variables. 
In this work, we closely follow \cite{aa1} in regard to the PCE surrogate model. Furthermore, choosing a polynomial order of $p=4$ and 60\% of the random sample for training matches the performance of the NN model most closely.

The \moo\ optimization with the surrogate models is done using the Python package DEAP~\cite{DEAP} and its standard implementation of NSGA-II. We picked hyperparameters that were as close as possible to those used with the OPAL GA.

\subsection{Code Availability}

For this research only open source software is used. This includes the accelerator simulation framework \opal~\cite{opal} and Python-based software tools: DEAP~\cite{DEAP}, Keras~\cite{chollet2015keras}, TensorFlow~\cite{tensorflow2015-whitepaper}, UQTk~\cite{Debusschere2016},  and sci-kit learn~\cite{scikit-learn}. 

\bibliography{surrogate-models-reformat}

\begin{addendum}
 \item  We gratefully acknowledge the computing resources provided on Bebop, a high-performance 
 computing cluster operated by the LCRC at ANL. The training of the surrogate models benefited from the ETH Leonard cluster the CSCS Piz-Daint, the PSI Merlin-6 and the SLAC OCIO Jupyter-Hub GPU cluster. We also thank John Power for assistance in developing the simulation model of the AWA (which was built by N. Neveu during her graduate studies with the AWA), and we acknowledge D. Winklehner and L. Calabretta in the development of the IsoDAR simulations. 
 
 This work was supported by the U.S. Department of Energy, Office of Science, under contract numbers DE-AC02-76SF00515, DE-AC02-06CH11357, and grant number DE-SC0015479. 
 
 \item[Competing Interests] The authors declare that they have no
competing financial interests.
\end{addendum}

\end{document}